\documentclass[aps,prl,twocolumn, showpacs, groupedaddress]{revtex4}

\usepackage{amsmath}
\usepackage{amssymb}
\usepackage{graphics}
\usepackage[dvips]{graphicx}
\graphicspath{{./}}
\begin{document}

\title{Vortex Ring Reconnections}

\author{Philippe Chatelain}
\author{Demosthenes Kivotides}
\author{Anthony Leonard}
\affiliation{Graduate Aeronautical Laboratories\\California Institute of Technology\\Pasadena, CA 91125}

\date{\today}

\begin{abstract}
We investigate numerically the Navier-Stokes dynamics
of reconnecting vortex rings at small $Re$ number. We find that 
reconnections are dissipative due to the smoothing
of vorticity gradients at reconnection kinks and to the 
formation of secondary structures of stretched anti-parallel vorticity
which transfer kinetic energy to small scales where it is 
subsequently dissipated efficiently. In addition, the relaxation of 
the reconnection kinks excites Kelvin waves
which due to strong damping are of low wavenumber and 
affect directly only large scale properties of the flow.
\end{abstract}


\maketitle

In flow phenomena as diverse as quantum \cite{europhys:kivotides:2002}
magnetic \cite{physrevD:christensson:1999} and incompressible \cite{annrevfm:leonard:1985} fluids,
it is useful to study the physics of turbulence
by modeling the system as a collection of tubular flux loops
which in the case of vortical fields are called vortex filaments.
An intrinsic property of such highly structured systems is their ability
to dynamically change their topology via reconnection mechanisms.
Does this change in topology affect in turn properties of fluid turbulence like the intermittency and
the scalar-mixing (which depend directly on the structure of the flow) or like
the dynamics of energy in wavenumber space? Or is it the case that reconnection
events are not generic and thus have no direct impact on the mean properties of
turbulent flows? The aim of this letter is to address
these issues by fully resolving the Navier-Stokes dynamics of interacting vortex rings for 
three simple geometries having great potential for illuminating the physics of reconnection. 
Although the flows considered are not strictly turbulent, the hope is that in a future structural
approach to the problem of turbulence a significant part of the flow complexity
could be traced back to the physics of similar vortex interactions.\\
Incompressible vortex reconnections have an extensive bibliography 
(for a review of the work up to 1994, 
see \cite{annrevfm:kida_takaoka:1994}).
In  \cite{jfm:shelley_meiron:1993,prl:pumir_kerr:1987} reconnections of
vortex tubes were considered with an emphasis on the possibility of
singularity formation as $Re \rightarrow \infty$.
In \cite{ctr:winckelmans:1995} the 
strong interactions between vortex rings were computed with the interest in 
developing numerical methods and turbulence models rather than in focusing on
the physics of reconnection. In \cite{nature:aref_zawadzki:1991} it is discussed
how a linked vortex configuration could be achieved starting from an unlinked
initial state and in  \cite{physfluidsa:zawadzki_aref:1991} it is considered 
how the mixing of a nondiffusing passive scalar is affected during vortex ring collision.
The reconnection of two approaching (but not colliding) vortex rings was studied 
experimentally in \cite{galcit:schatzle:1987} and theoretically in \cite{prl:ashurst_meiron:1987}.
This letter extends these studies by considering generic vortex configurations 
and by capturing more features of vortex reconnections in a turbulent flow.

We solve the Navier-Stokes equations for an unbounded three-dimensional incompressible viscous flow. We employ the vorticity formulation:
\begin{equation}
\label{eq_vorticity}
\left(\frac{\partial}{\partial t}+\boldsymbol{u}\cdot\nabla\right)\boldsymbol{\omega} = \left(\nabla\boldsymbol{u}\right) \cdot \boldsymbol{\omega} + \nu \nabla^2 \boldsymbol{\omega}\;,
\end{equation}
\begin{equation}
\nabla \cdot \boldsymbol{u} = 0\;,
\end{equation}
where $\boldsymbol{u}$ is the velocity and $\boldsymbol{\omega}$ is the vorticity.
We use a vortex particle method \cite{cottet_koumoutsakos}. In this method, the vorticity is discretized with Lagrangian elements.
These elements which carry a vector-valued Gaussian distribution of vorticity 
are convected and stretched by the local velocity 
obtained by the Biot-Savart law. The complexity of the velocity computation 
is normally $\mathcal{O}(N^2)$ with $N$ being the number of particles; we have used a multipole 
algorithm that reduces this complexity to $\mathcal{O}(N \log(N))$. The viscous diffusion is modeled 
by the Particle Strength Exchange scheme.\\
We calculate some global quantities: the kinetic energy $E$, the enstrophy $\Omega$ and the helicity $H$, defined as
\begin{eqnarray}
E & = & \frac{1}{2}\int \boldsymbol{u}\cdot\boldsymbol{u}\, d\boldsymbol{x}\,,\\
\Omega & = & \int \boldsymbol{\omega}\cdot\boldsymbol{\omega}\, d\boldsymbol{x}\,,\\
H & = & \int \boldsymbol{\omega}\cdot\boldsymbol{u}\, d\boldsymbol{x}\,.
\end{eqnarray}
For unbounded flows, the relation between kinetic energy and enstrophy is
\begin{equation}
\frac{d}{dt}E = -\nu \Omega\;.
\end{equation}

\begin{figure*}[t]
\begin{minipage}[t]{0.49\linewidth}
\begin{tabular}[b]{|c|c|c|c|}
\hline
\includegraphics[width=0.23\linewidth]{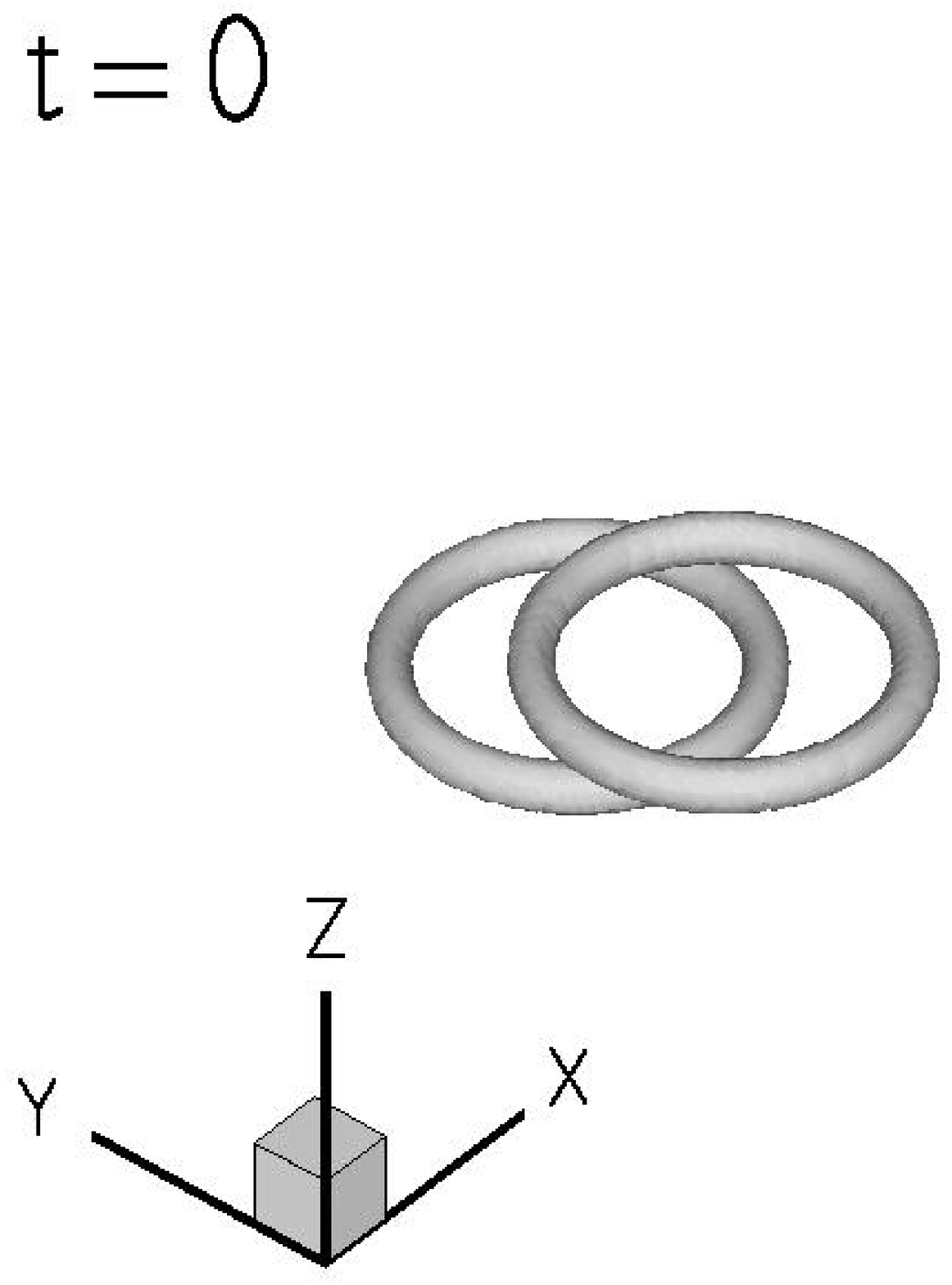} &
\includegraphics[width=0.23\linewidth]{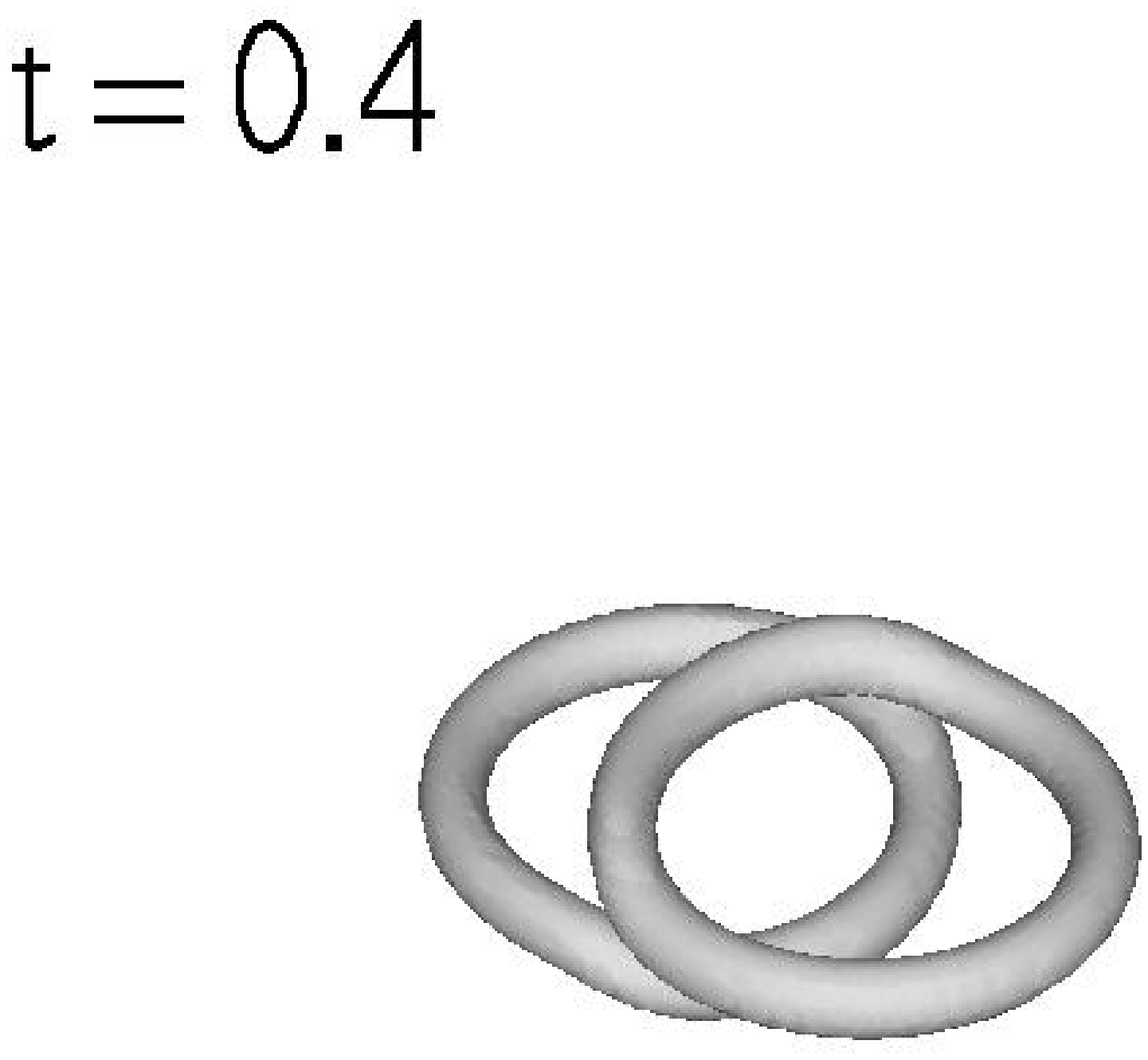} &
\includegraphics[width=0.23\linewidth]{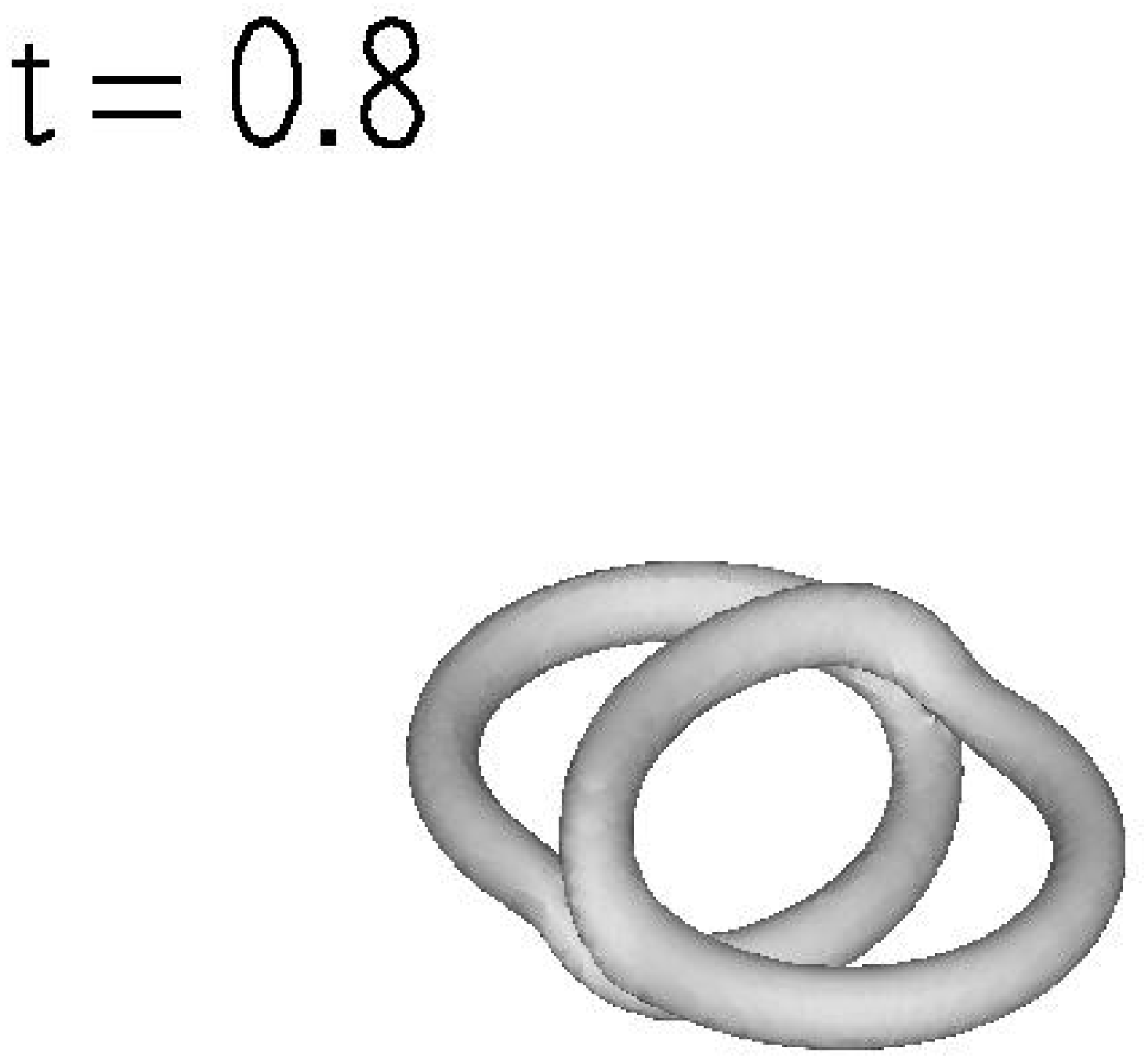} &
\includegraphics[width=0.23\linewidth]{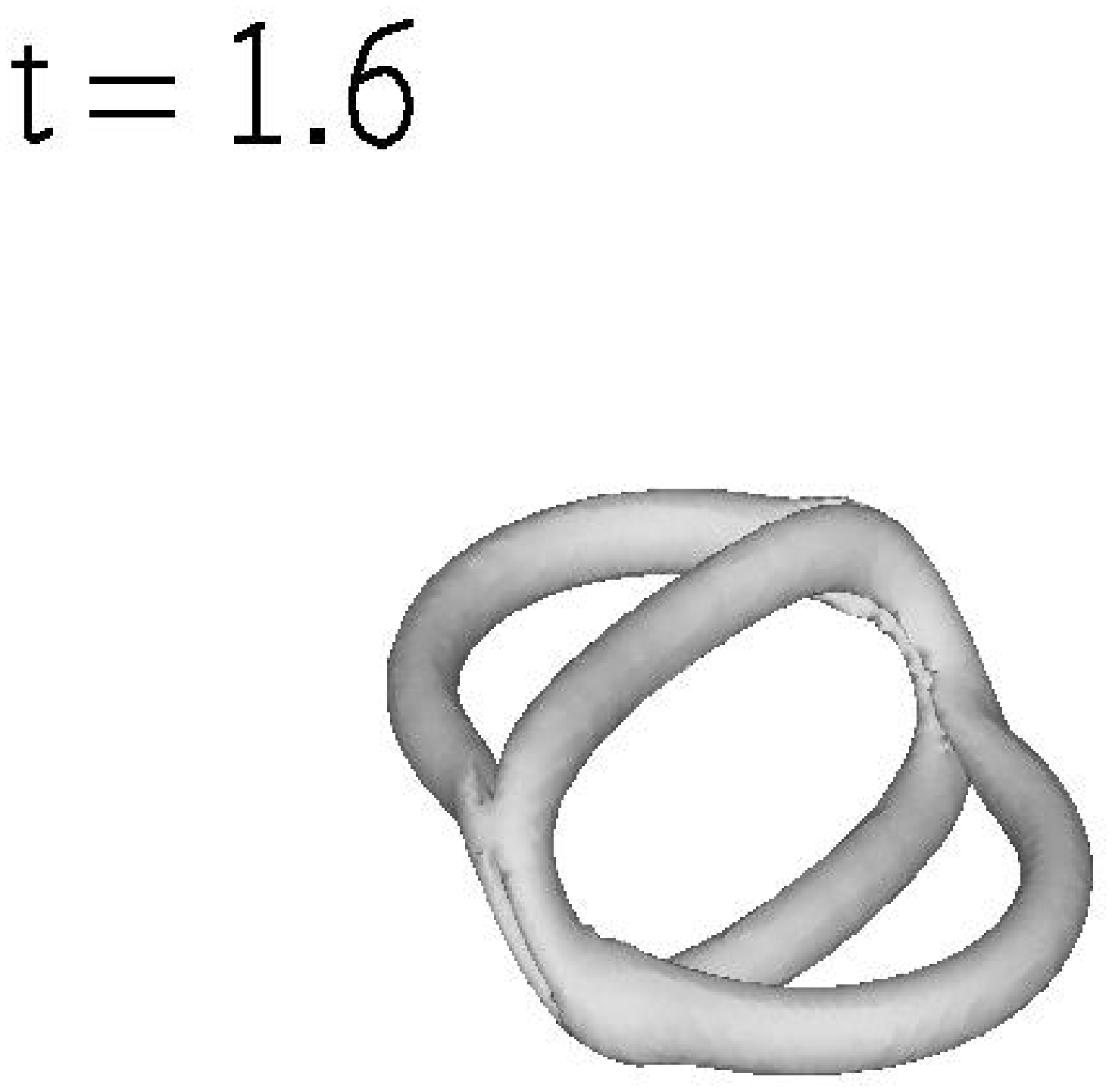}\\
\hline
\includegraphics[width=0.23\linewidth]{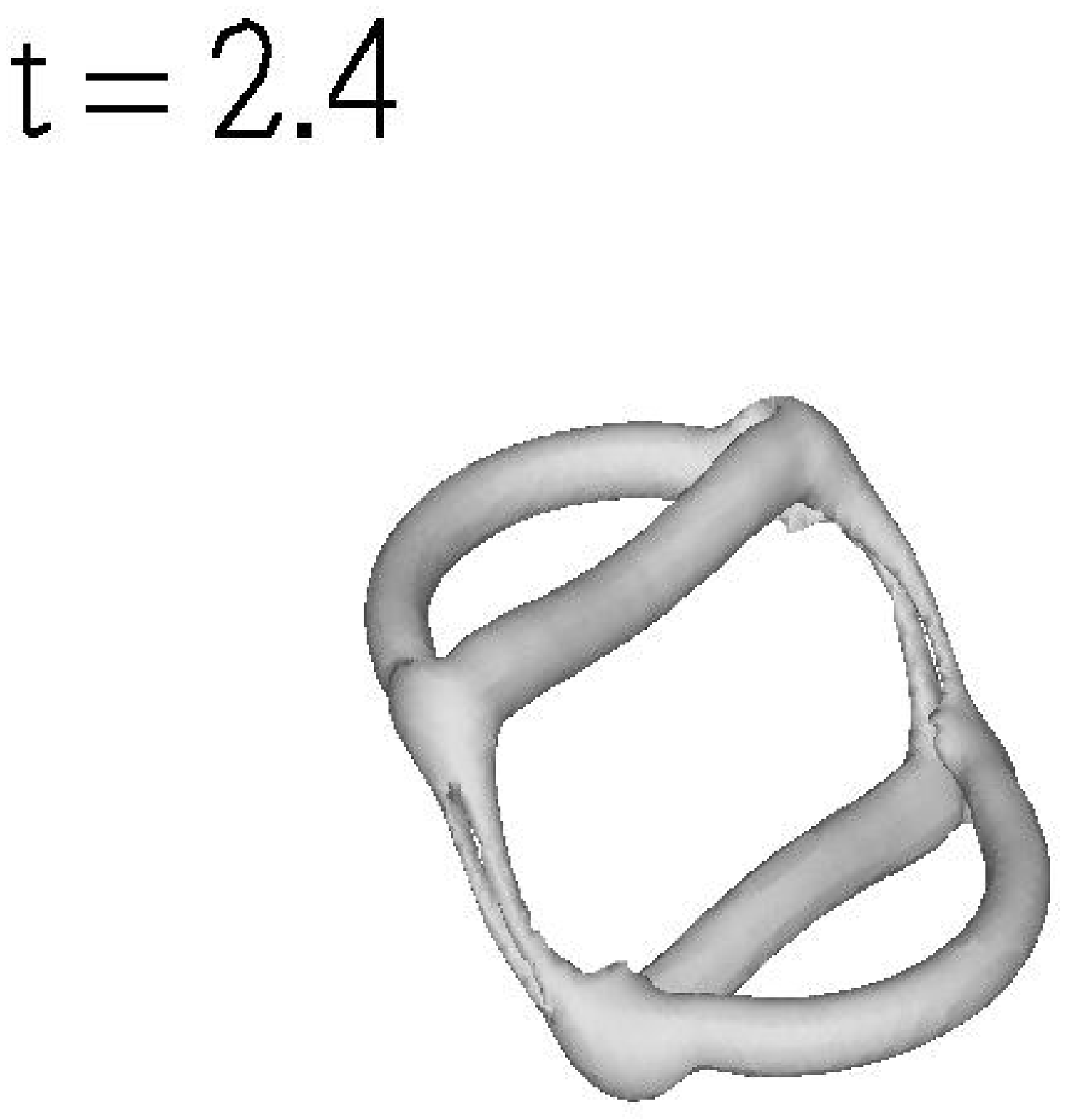} &
\includegraphics[width=0.23\linewidth]{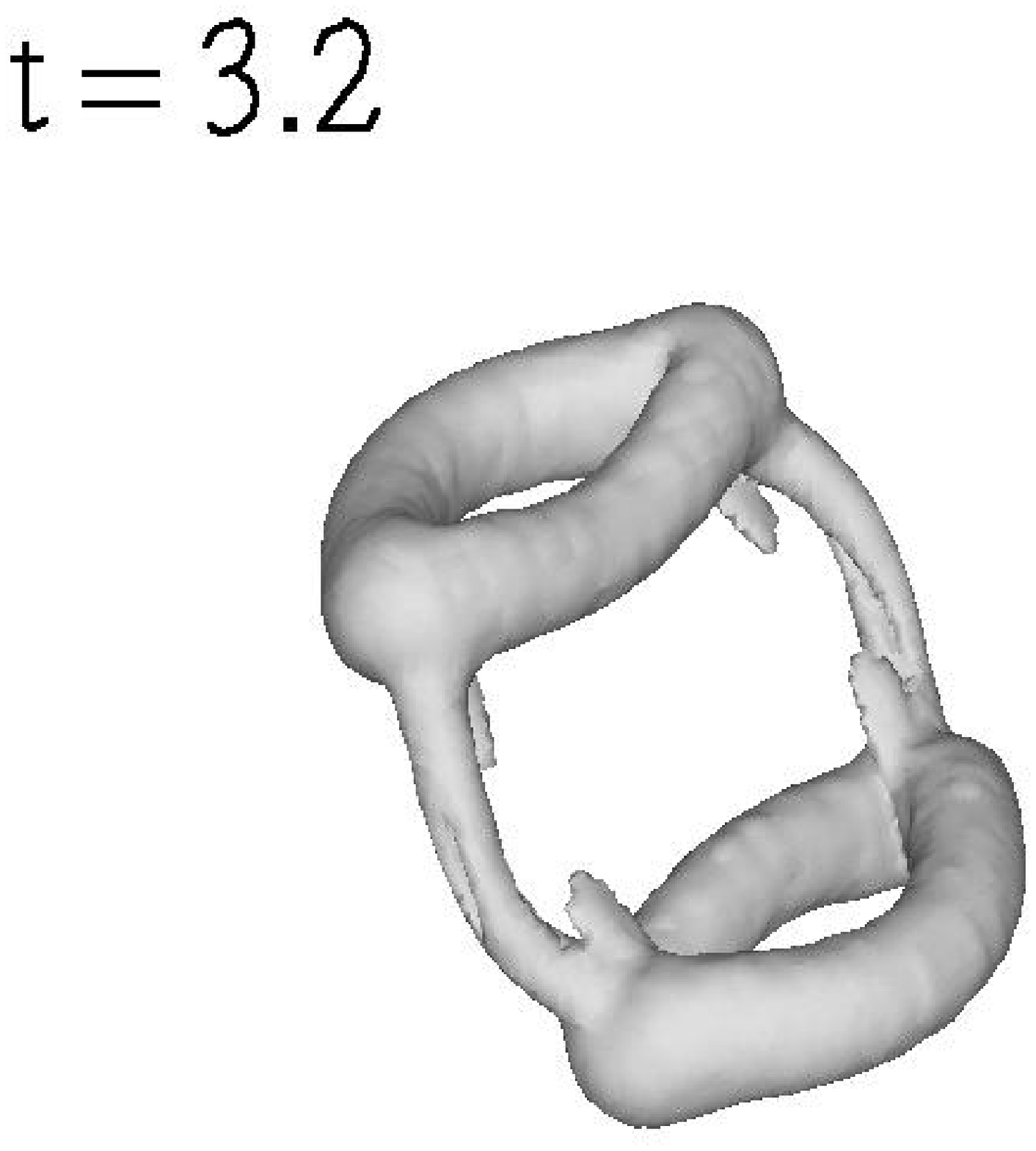} &
\includegraphics[width=0.23\linewidth]{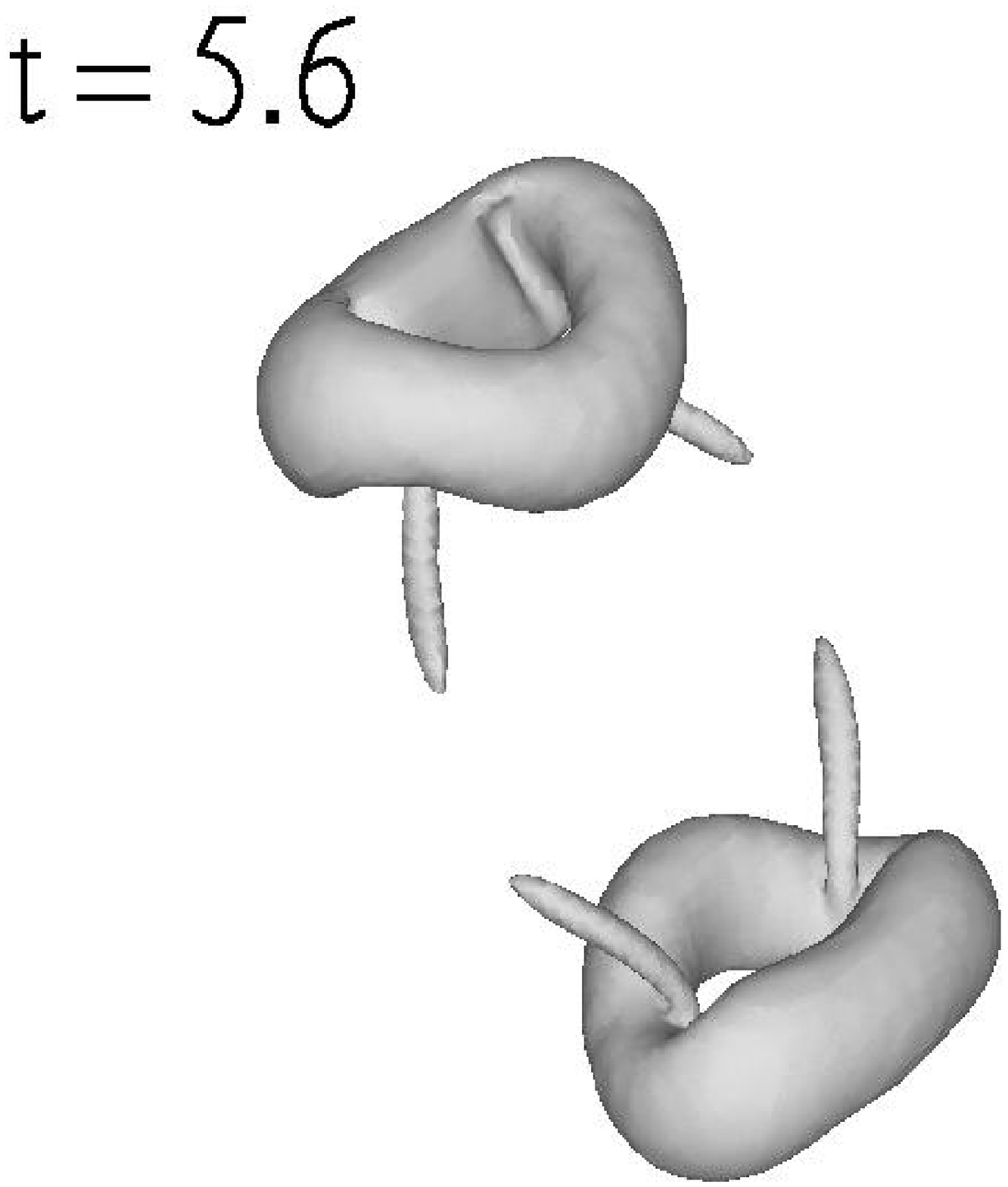} &
\includegraphics[width=0.23\linewidth]{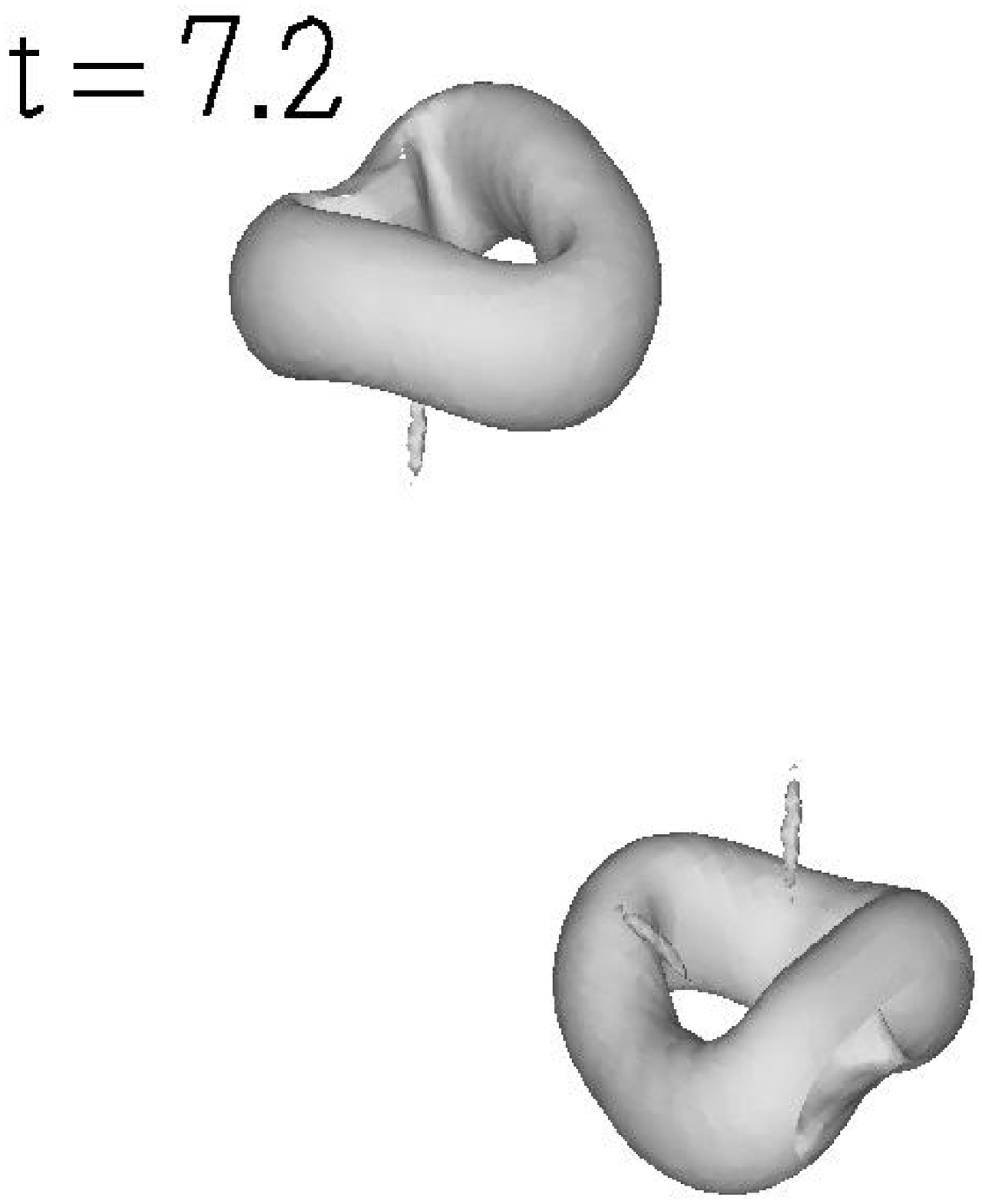}\\
\hline 
\end{tabular}
\caption{\label{figure:offcoll:evol_contours}Vortex rings in an offset collision: contours of vorticity; from $t=0$ to $2.4$, the contour is $\omega = 0.15\,\omega_{\text{max}}^{\text{t=0}}$, for $t>2.4$, it is $\omega = 0.025\,\omega_{\text{max}}^{\text{t=0}}$}
\end{minipage}
\hfill 
\begin{minipage}[t]{0.49\linewidth}
\includegraphics[width=\linewidth]{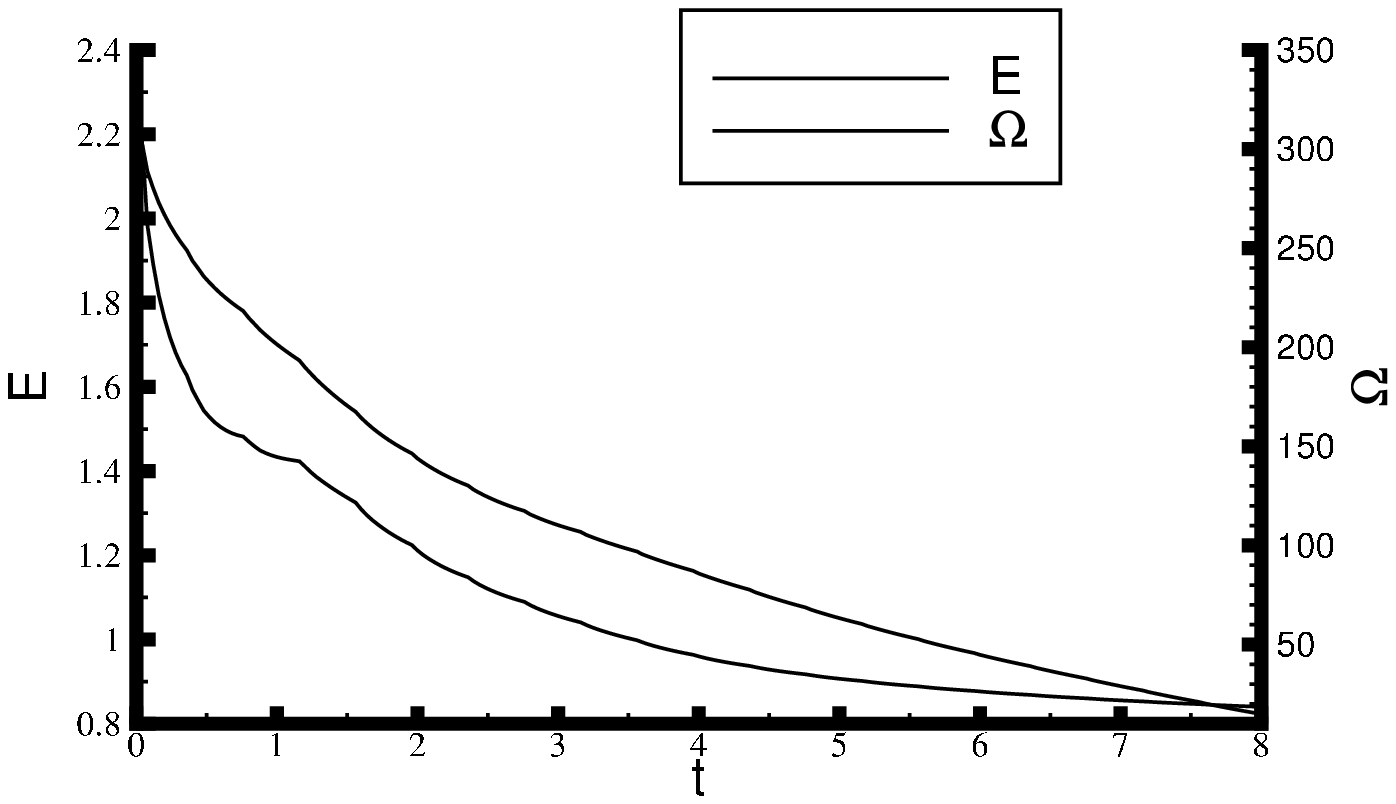}
\caption{\label{figure:offcoll:evol_stats}Vortex rings in an offset collision: kinetic energy and enstrophy}
\end{minipage}
\end{figure*}
We also compute the evolution of the spectrum of the kinetic energy $E(k)$ which, in terms of the Fourier transform of vorticity $\hat{\boldsymbol{\omega}}= \frac{1}{(2\pi)^{3/2}}\int \boldsymbol{\omega}(\boldsymbol{r})\,e^{-i\,\boldsymbol{r}\cdot\boldsymbol{k}}\,d\boldsymbol{r}$, is defined as
\begin{equation}
E(k) = \frac{1}{2} (2\pi)^3 \int_{|\boldsymbol{k}|=k} \hat{\boldsymbol{\omega}}\cdot\hat{\boldsymbol{\omega}}^\ast\,d\Omega_k\;,
\end{equation}
where $d\Omega_k$ denotes $\sin\theta_k\,d\theta_k\,d\phi_k$, the solid angle element in spherical coordinates.
The calculation of the spectrum requires a double summation over the vortex elements which results 
to $\mathcal{O}(N^2)$ complexity. Because of this, the calculation of the spectrum is much more costly than the solution of the Biot-Savart law. Since the number of particles grows substantially during our simulations, from around $N=5\,10^4$ at $t=0$ to $8\,10^5$ in the end, our computational resources did not allow us to compute the spectra for all times.

We consider three reconnection configurations at the same Reynolds number: $Re = \frac{\Gamma}{\nu} = 250$ 
where $\Gamma$ is the circulation of one ring and $\nu$ is the kinematic viscosity. This small value 
of the $Re$ was dictated by the computational cost and the need for well-resolved 
reconnection regions. All the rings have the same initial $\Gamma$. 
All of our conclusions are conditioned upon the small
value of the  $Re$ number, as well as, on the common initial circulation
and should not be extrapolated uncritically to other settings.
The vorticity distribution in the cross-section of every ring is Gaussian with a cut-off
\begin{equation}
\omega_\theta = \frac{\Gamma}{2 \pi \sigma^2} e^{\frac{-r^2}{2\sigma^2}}
\end{equation}
where $r$ is the distance to the core center, $\sigma$ is the core radius and $\omega_\theta$ is the azimuthal vorticity. We chose $\sigma = 0.05\,R$ (where $R$ is the radius of the ring), to ensure that the rings are still thin when reconnections occur. Our results were made dimensionless in the following manner:
$t = \frac{\Gamma\,t'}{R^2}$, $x = \frac{x'}{R}$, $\omega = \frac{R^2\,\omega'}{\Gamma}$
where $t'$, $x'$, $\omega'$ are dimensional.

\begin{figure}[h]
\includegraphics[width=0.95\linewidth]{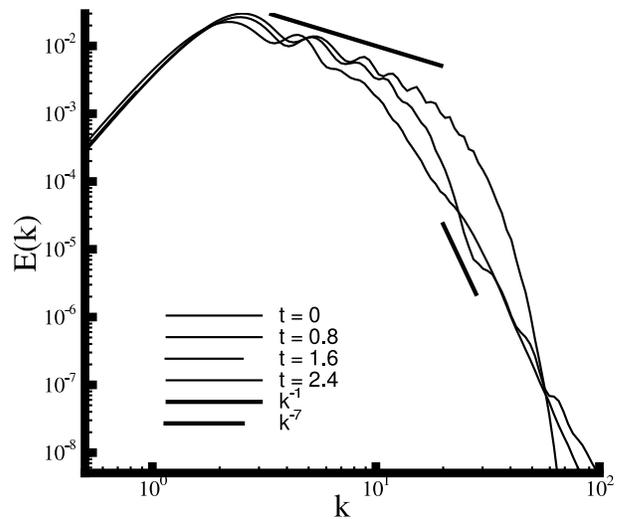}
\caption{\label{figure:offcoll:spectrum_evolution}Vortex rings in an offset collision: evolution of the energy spectrum}
\end{figure}
\begin{figure*}[t]
\begin{minipage}[t]{0.48\linewidth}
\begin{tabular}[b]{|c|c|c|c|}
\hline
\includegraphics[width=0.23\linewidth]{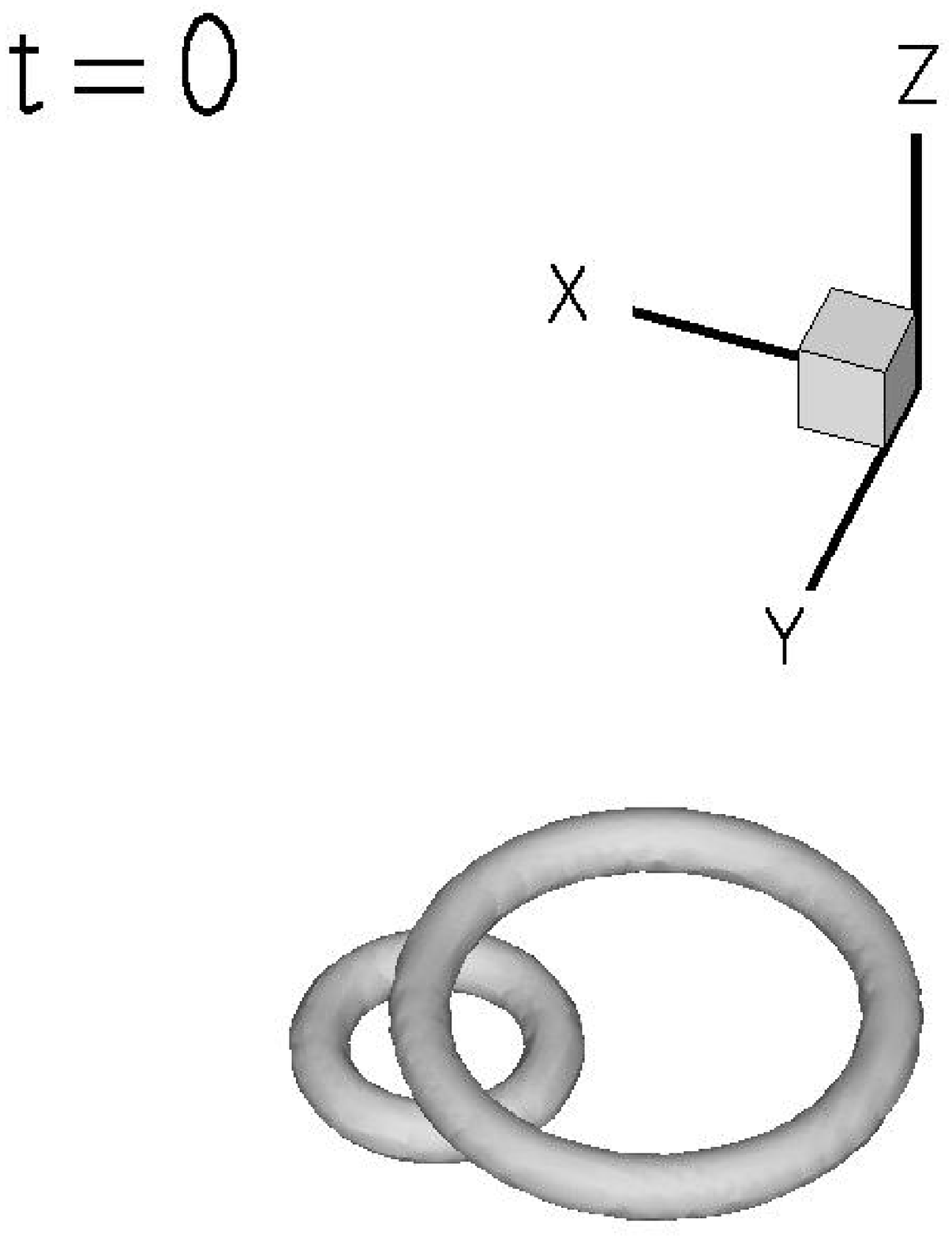} &
\includegraphics[width=0.23\linewidth]{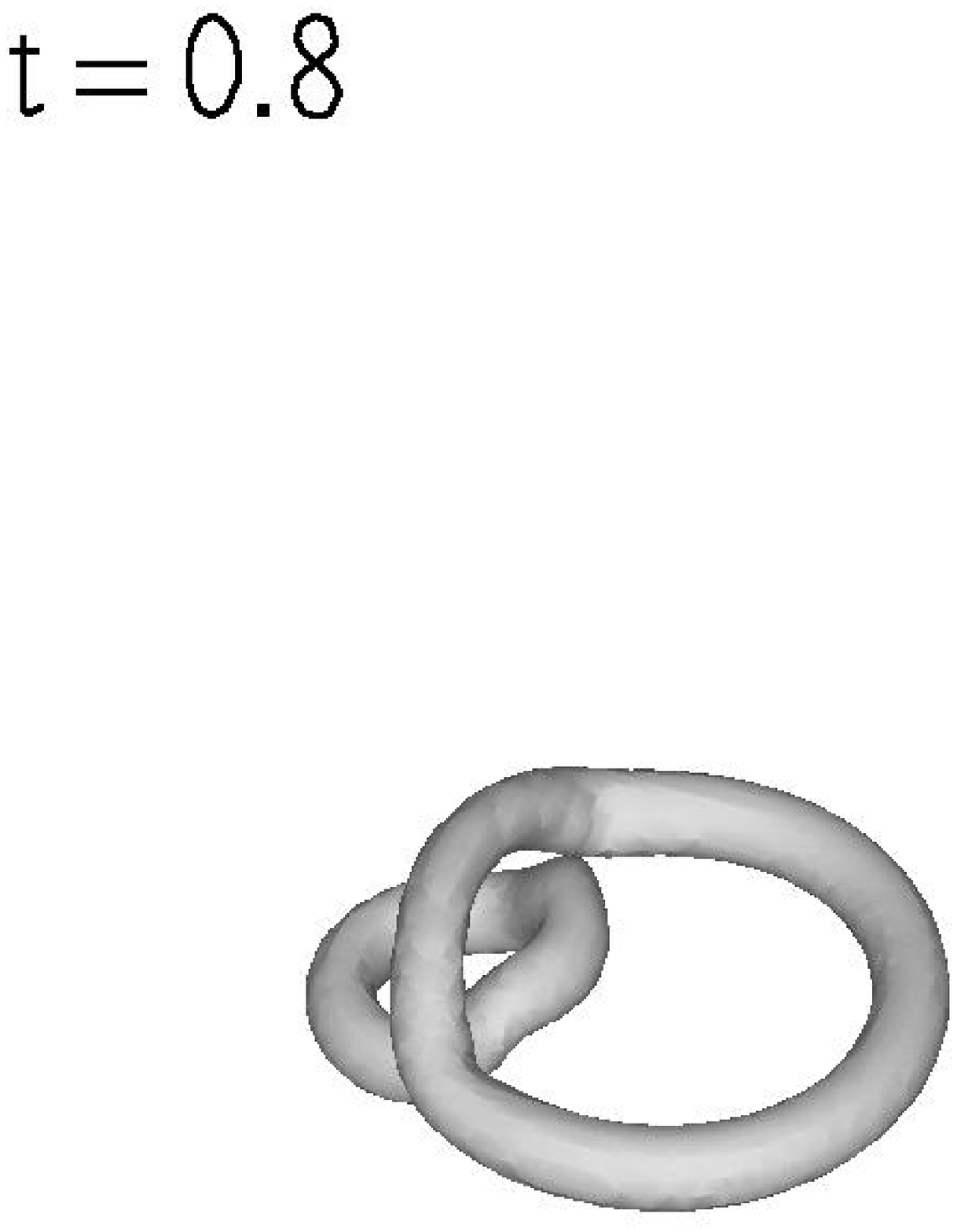} &
\includegraphics[width=0.23\linewidth]{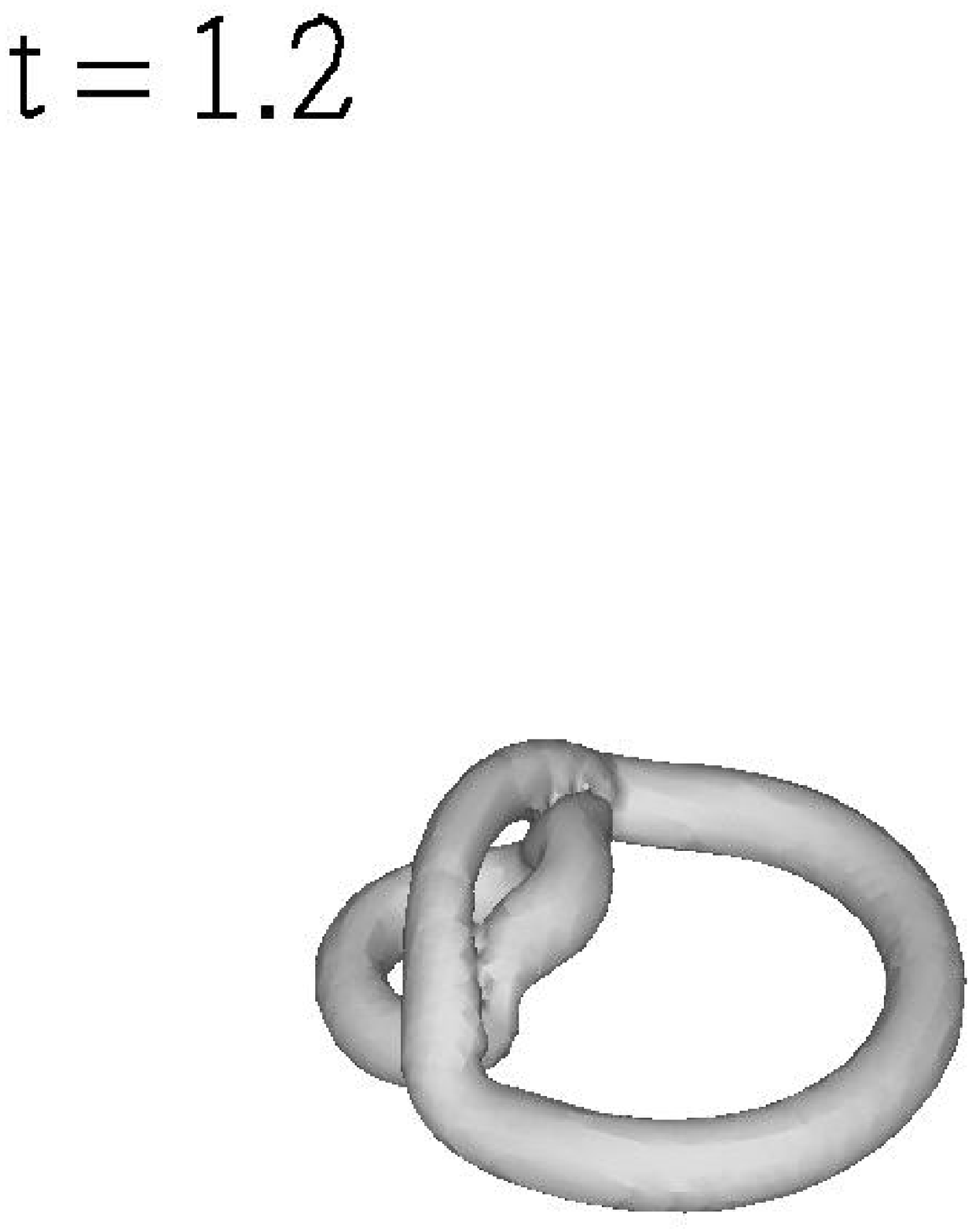} &
\includegraphics[width=0.23\linewidth]{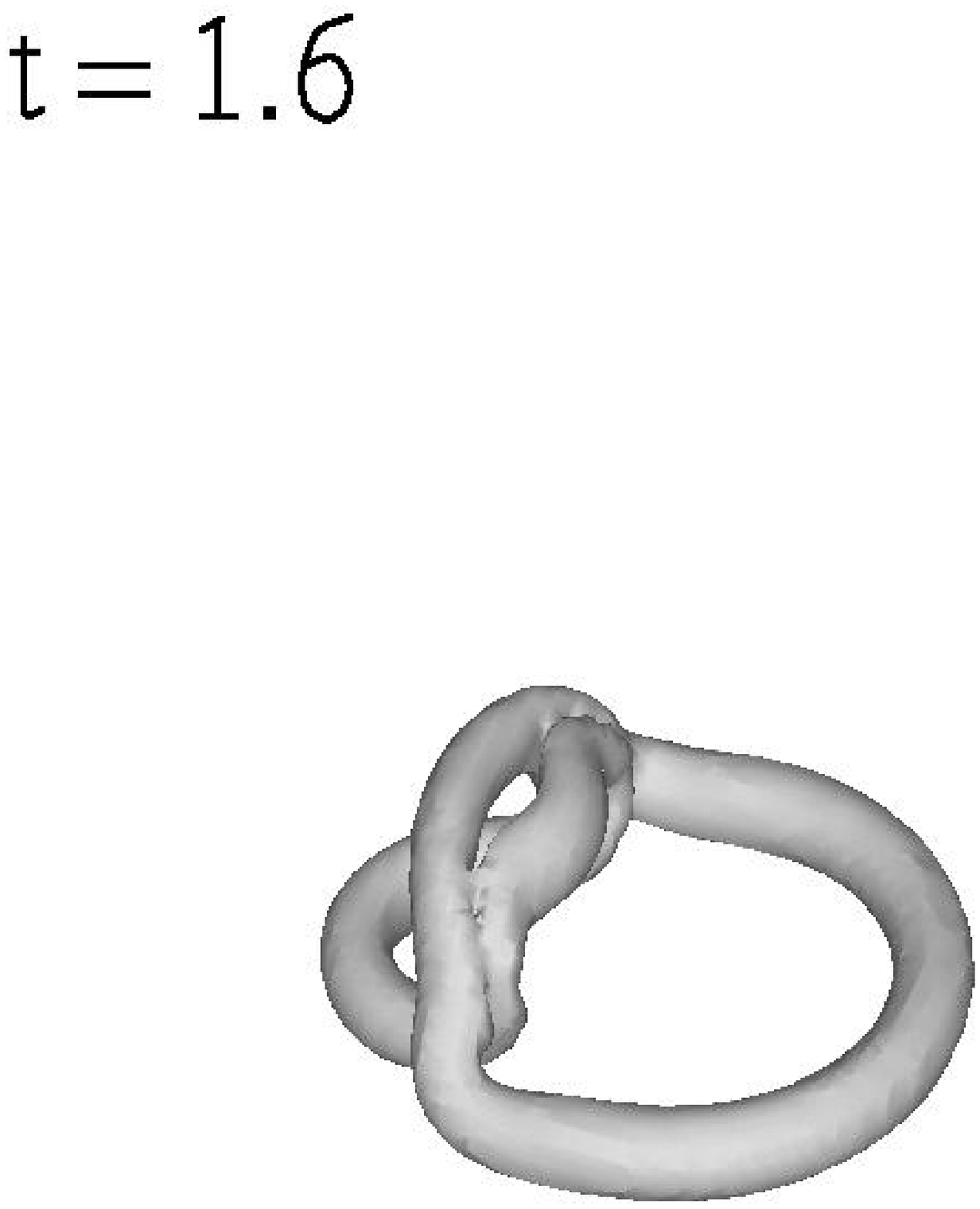}\\
\hline
\includegraphics[width=0.23\linewidth]{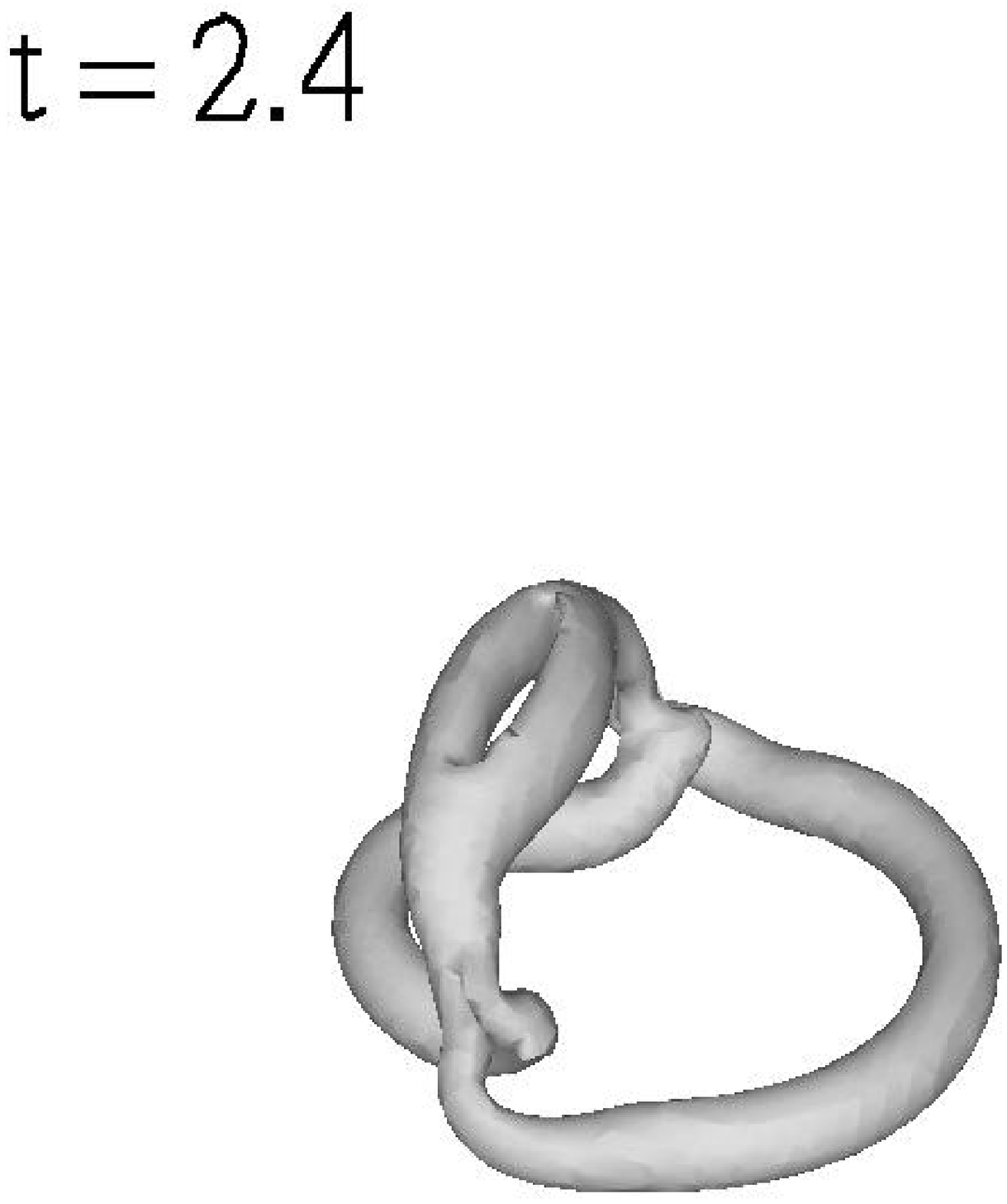} &
\includegraphics[width=0.23\linewidth]{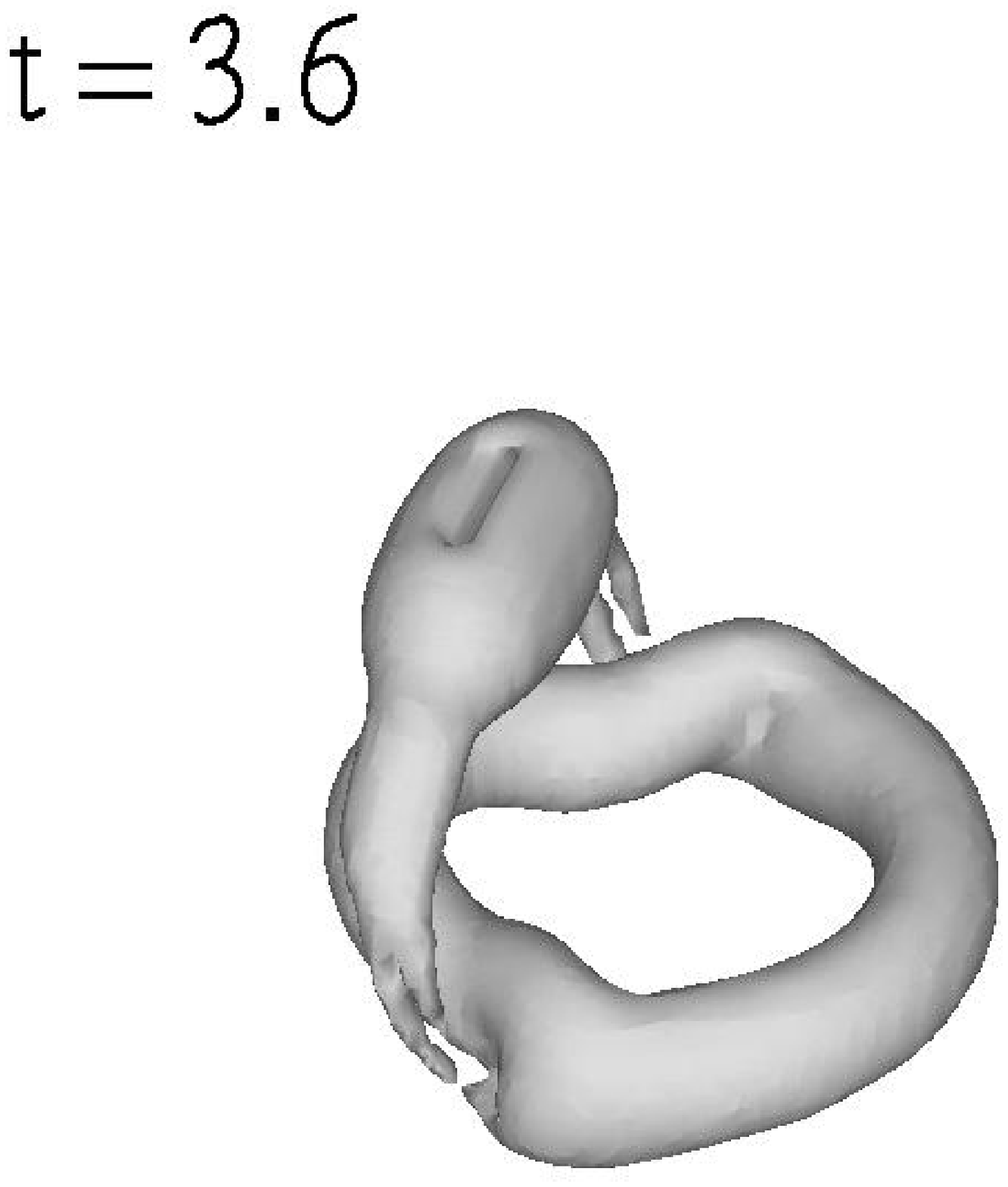} &
\includegraphics[width=0.23\linewidth]{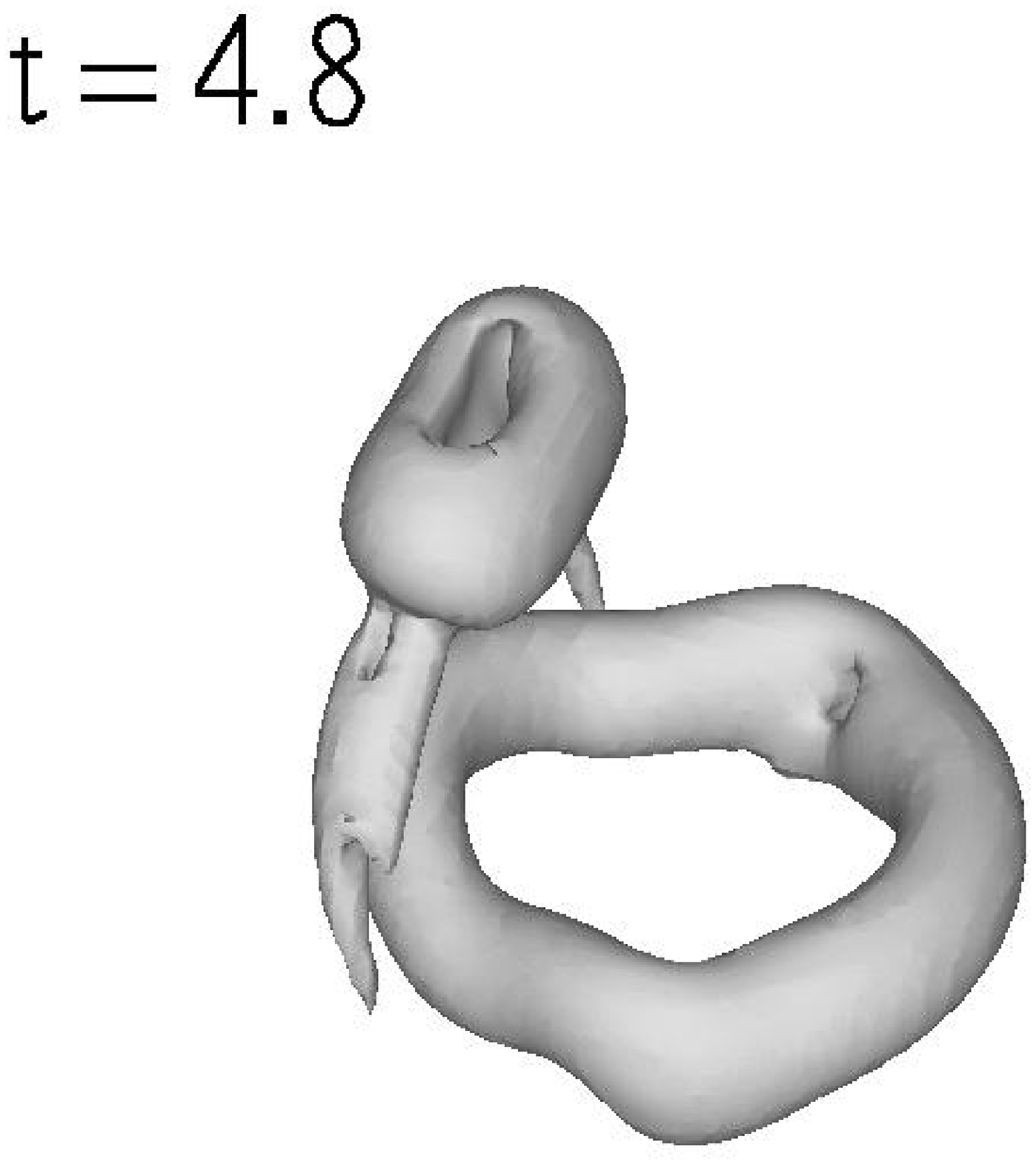} &
\includegraphics[width=0.23\linewidth]{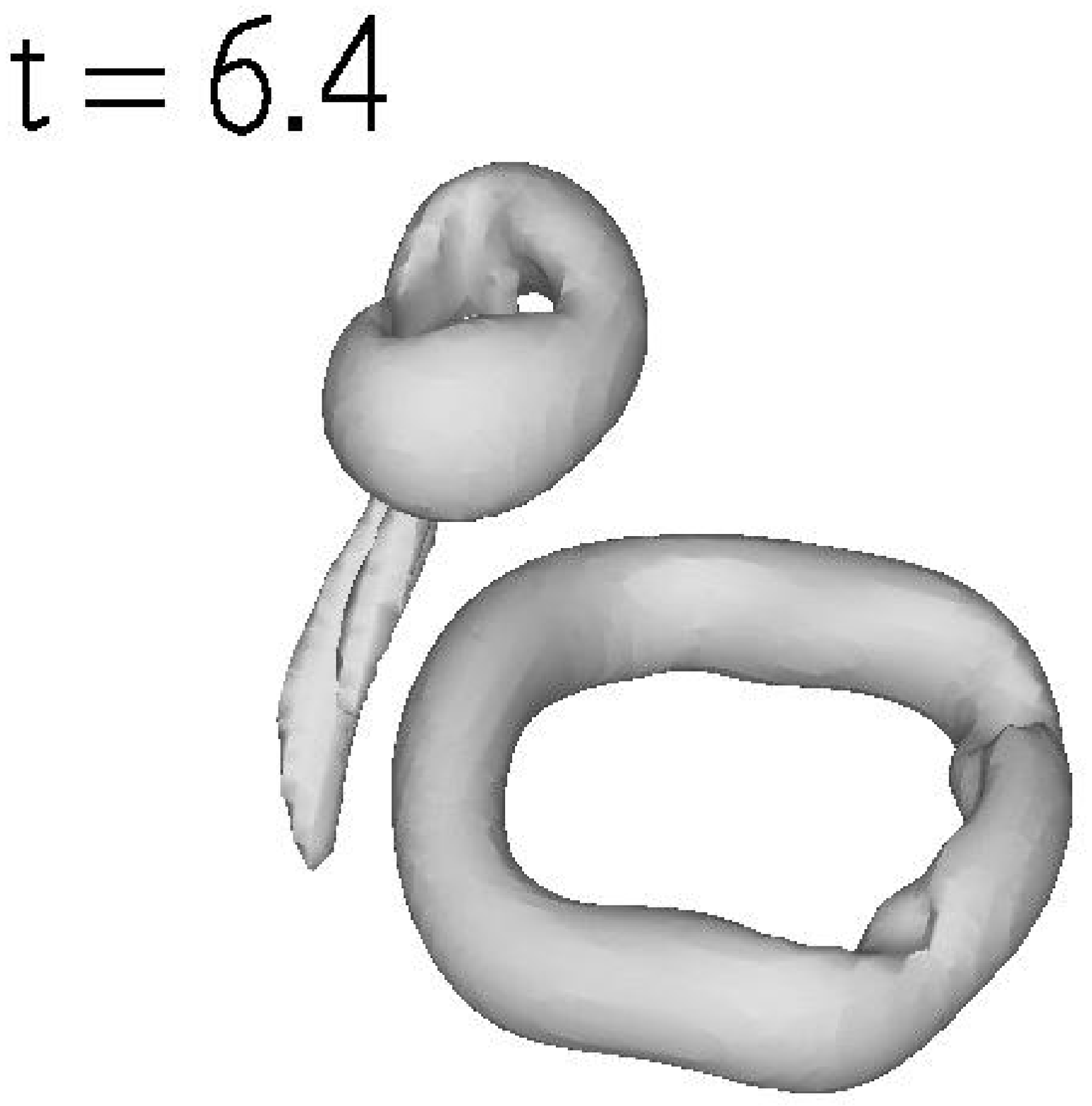}\\
\hline 
\end{tabular}
\caption{\label{figure:diffr:evol_contours}Vortex rings of different radii: contours of vorticity; from $t=0$ to $2.4$, the contour is $\omega = 0.15\,\omega_{\text{max}}^{\text{t=0}}$, for $t>2.4$, it is $\omega = 0.05\,\omega_{\text{max}}^{\text{t=0}}$}
\end{minipage}
\hfill 
\begin{minipage}[t]{0.48\linewidth}
\includegraphics[width=\linewidth]{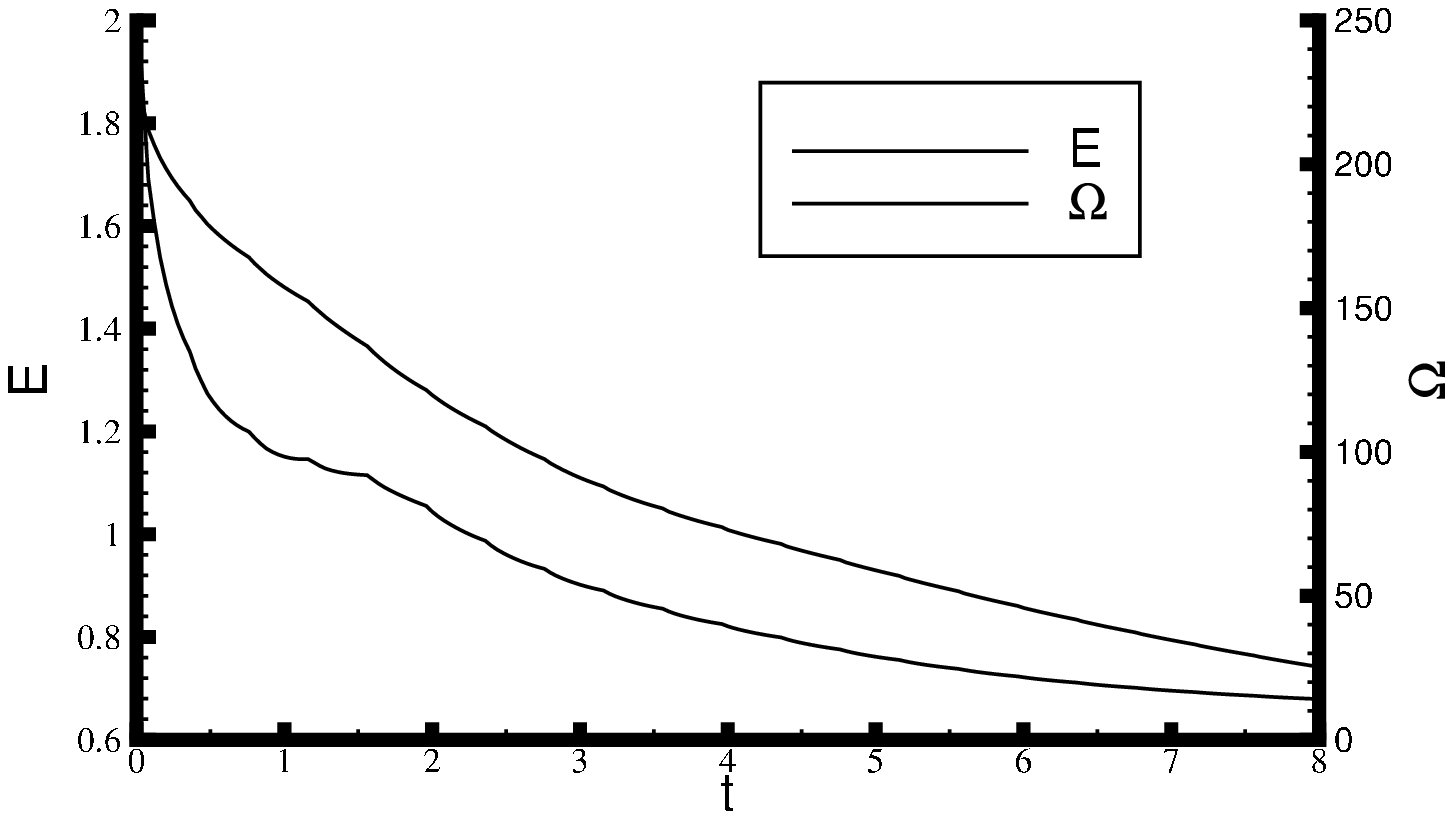}
\caption{\label{figure:diffr:evol_stats}Vortex rings of different radii: kinetic energy and enstrophy}
\end{minipage}
\end{figure*}
\begin{figure}[h]
\includegraphics[width=0.95\linewidth]{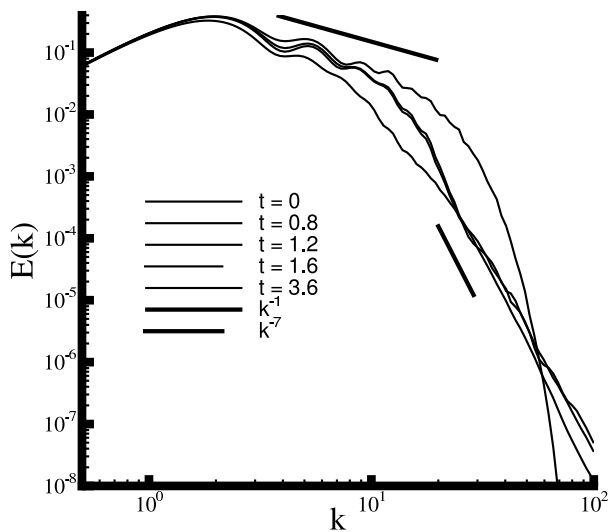}
\caption{\label{figure:diffr:spectrum_evolution}Vortex rings of different radii: evolution of the energy spectrum}
\end{figure}
\begin{figure*}[t]
\begin{minipage}[t]{0.48\linewidth}
\begin{tabular}[b]{|c|c|c|c|}
\hline
\includegraphics[width=0.23\linewidth]{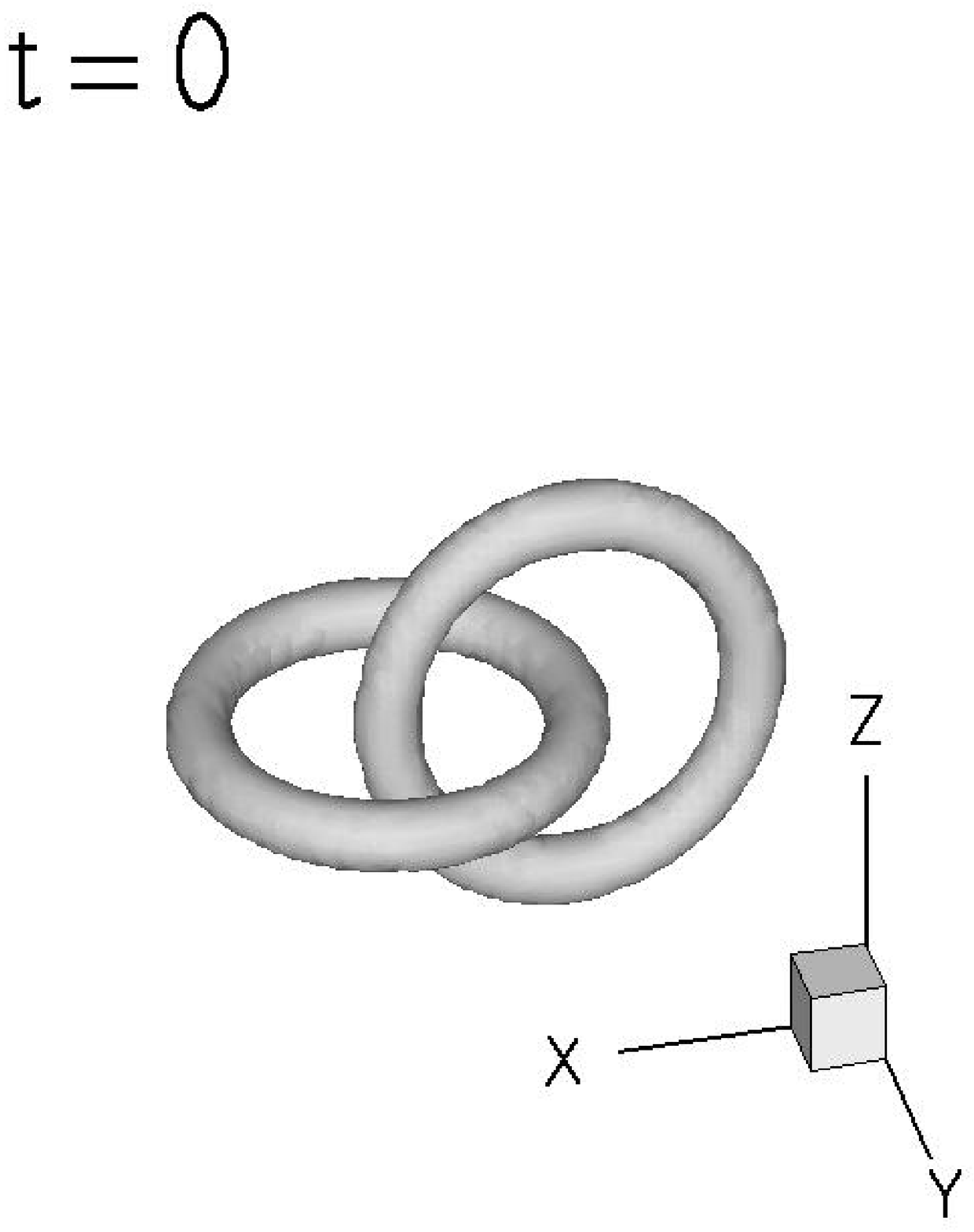} &
\includegraphics[width=0.23\linewidth]{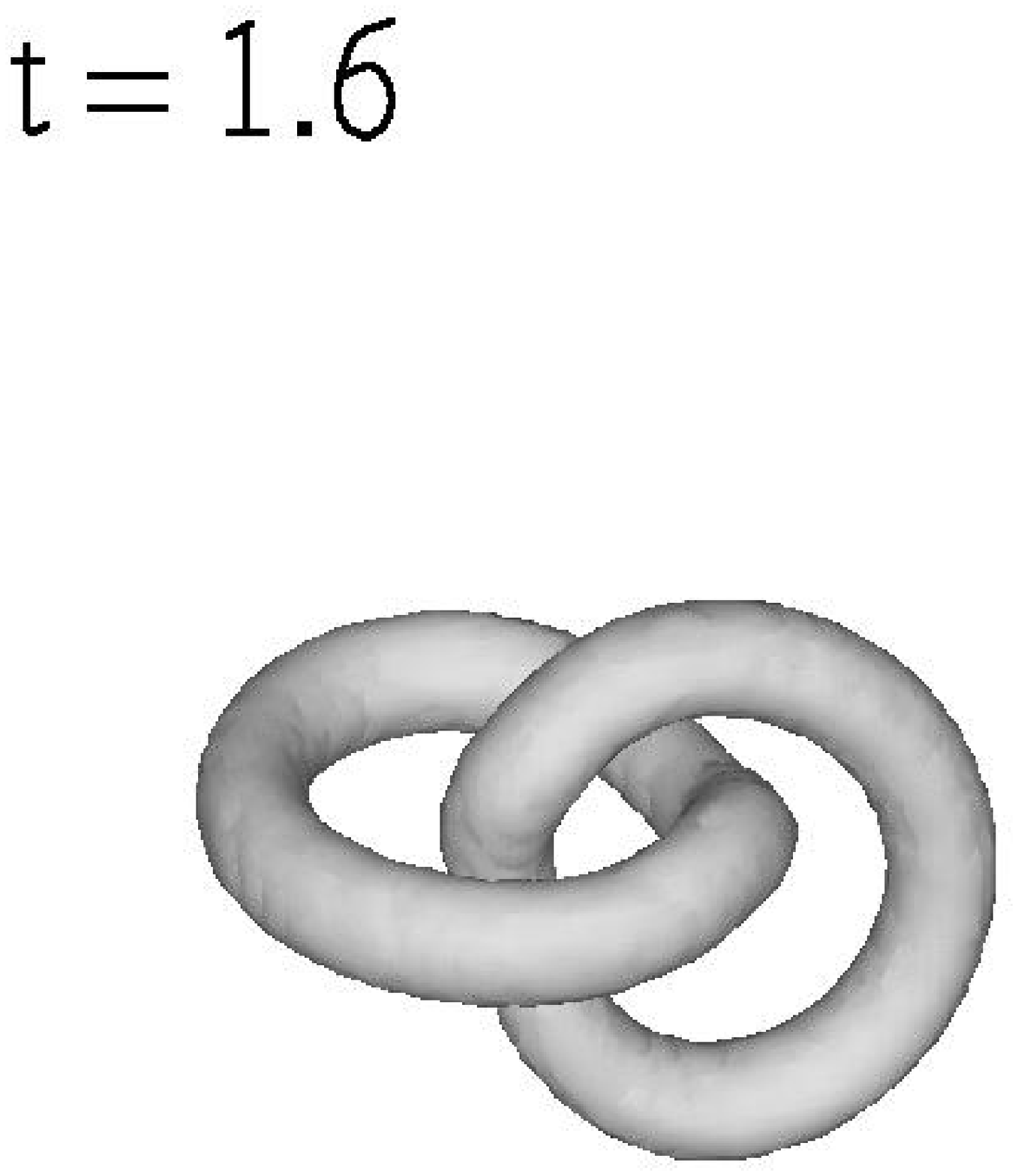} & 
\includegraphics[width=0.23\linewidth]{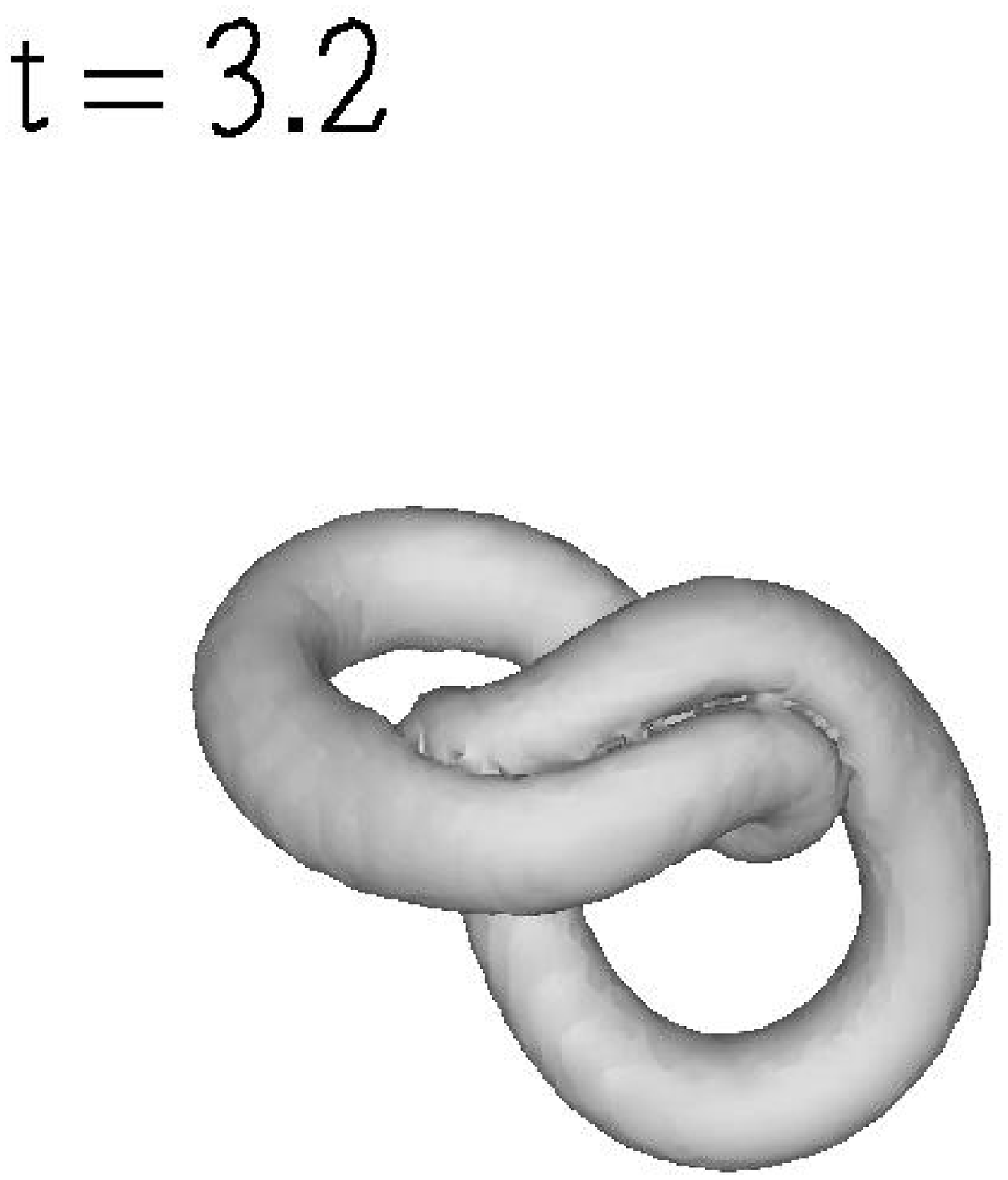} & 
\includegraphics[width=0.23\linewidth]{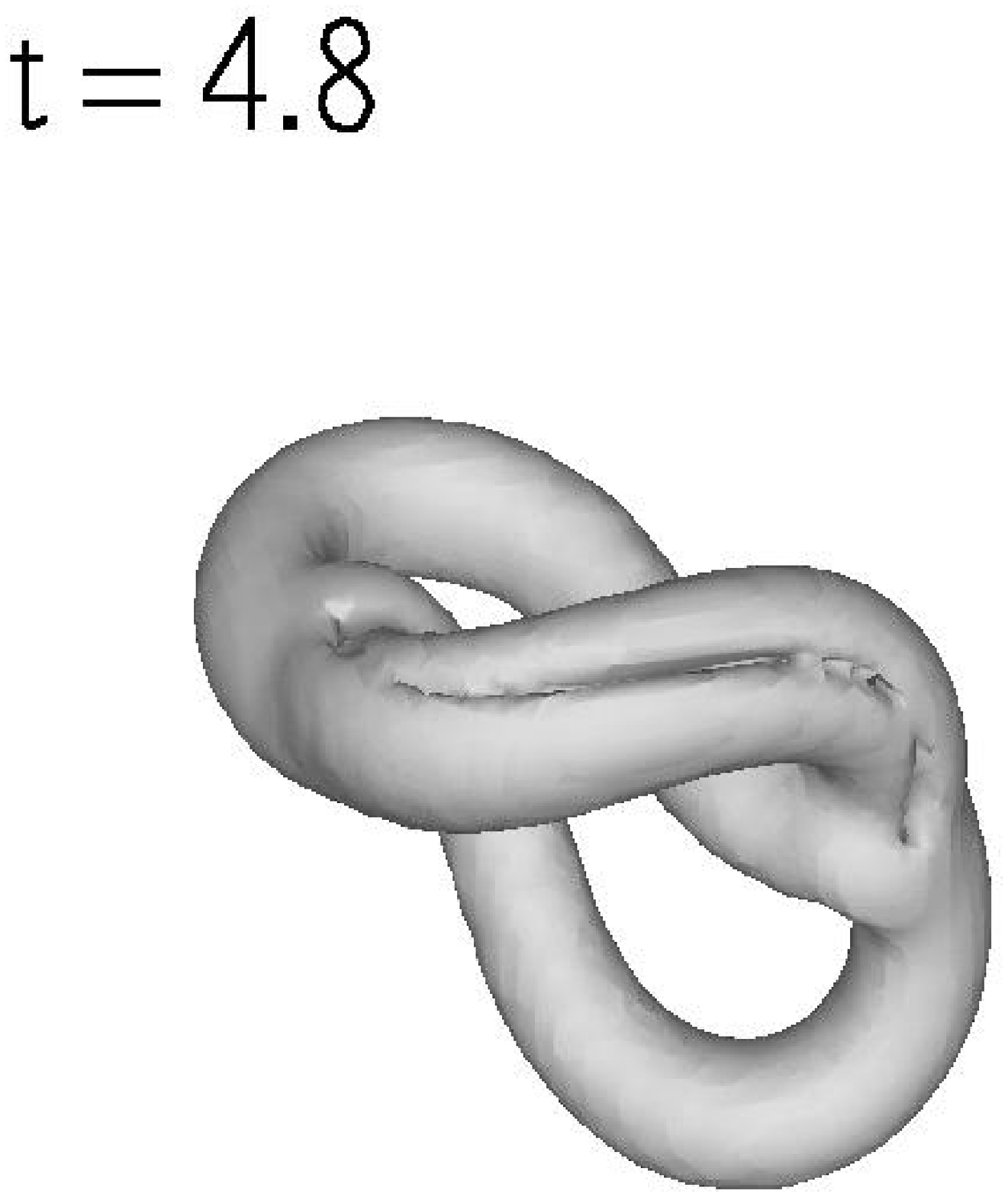}\\ 
\hline 
\includegraphics[width=0.23\linewidth]{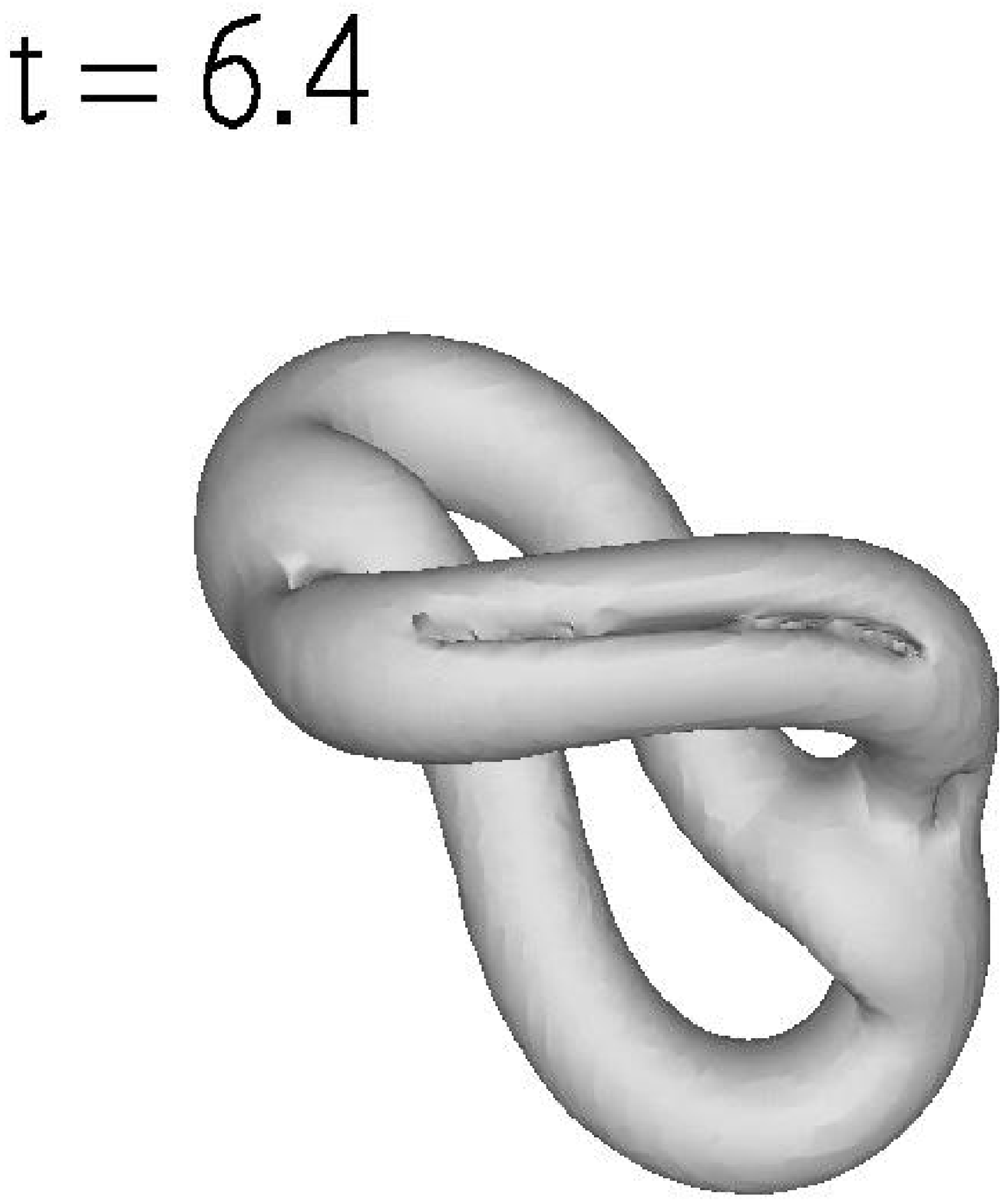} & 
\includegraphics[width=0.23\linewidth]{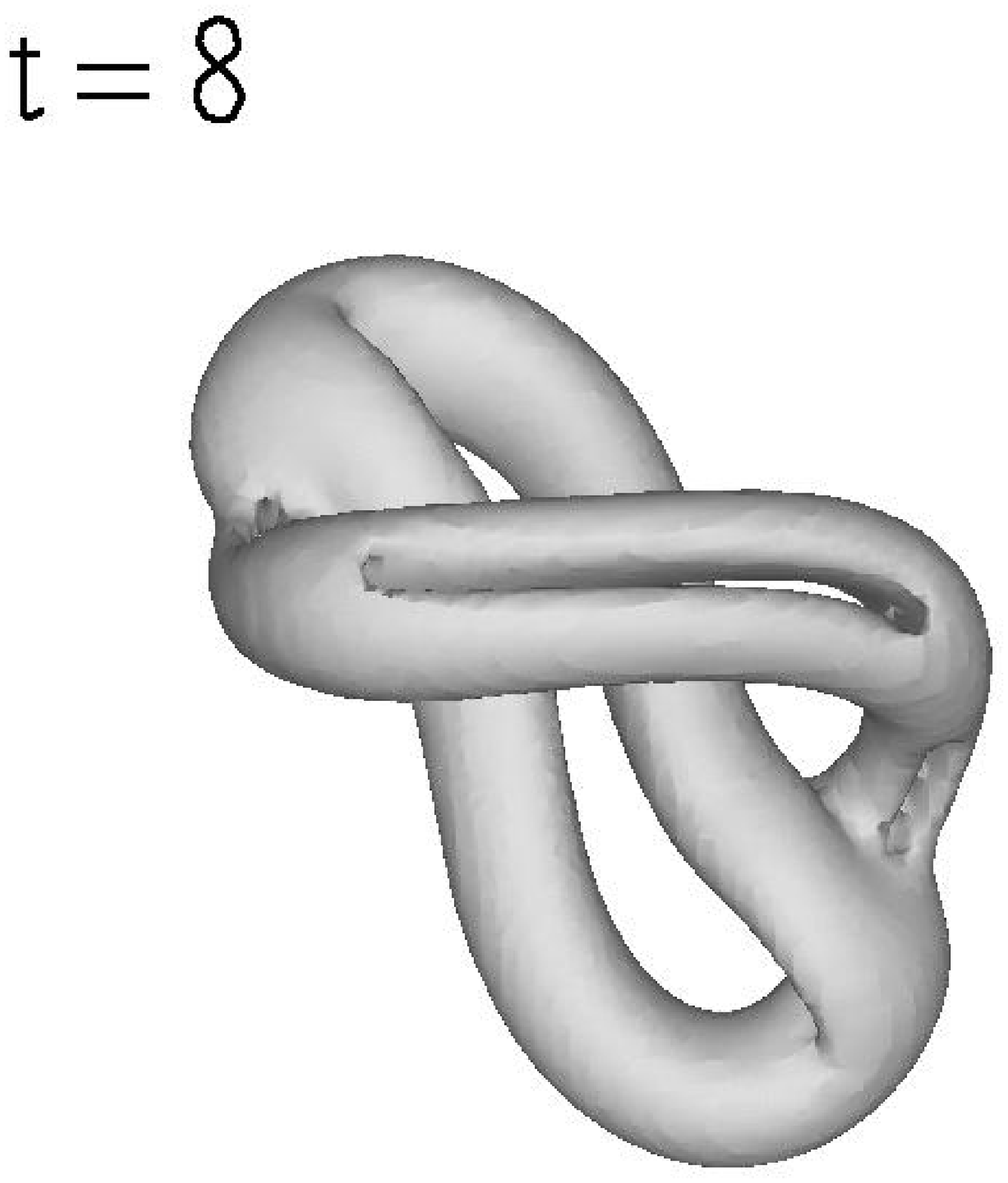} & 
\includegraphics[width=0.23\linewidth]{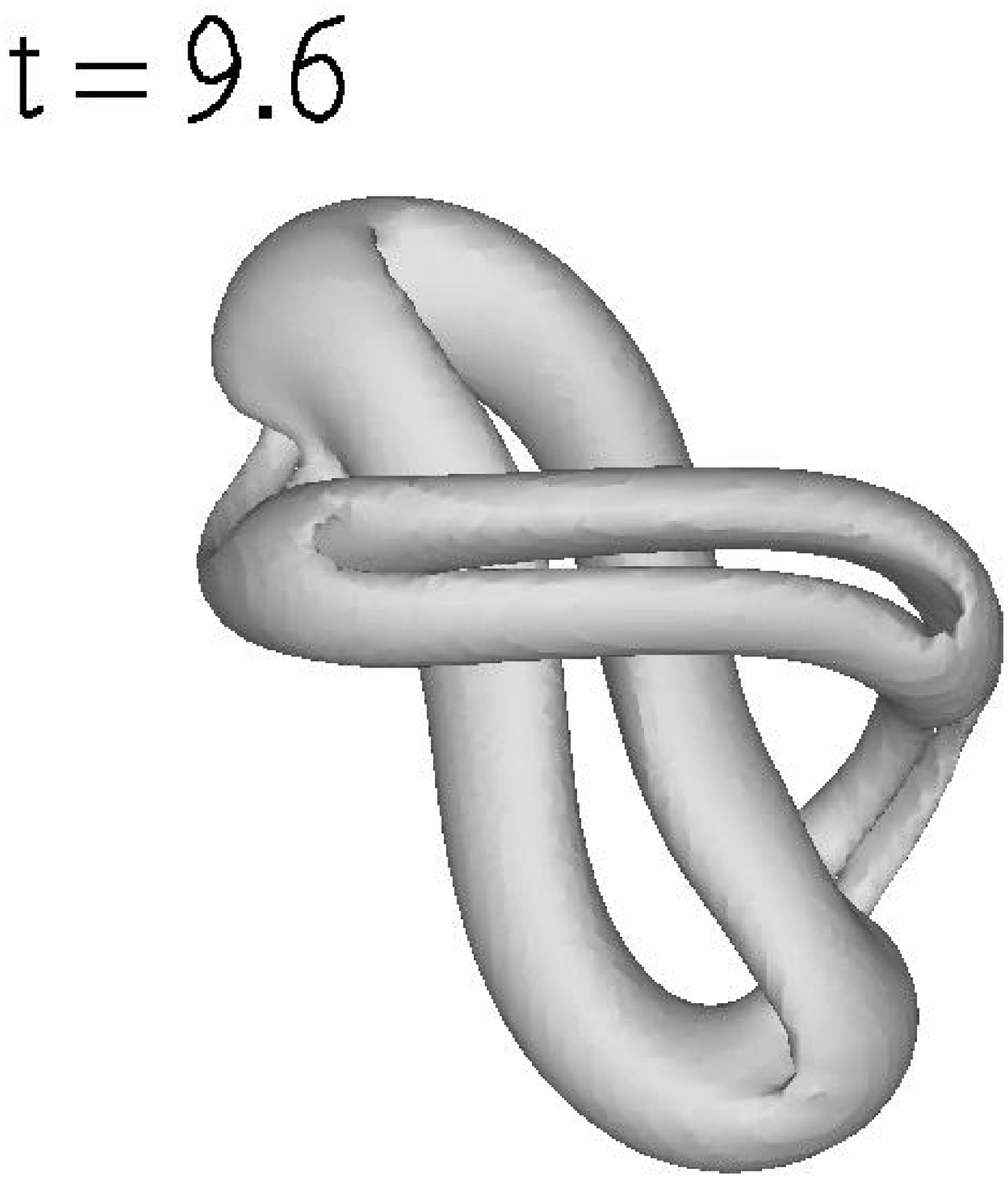} & 
\includegraphics[width=0.23\linewidth]{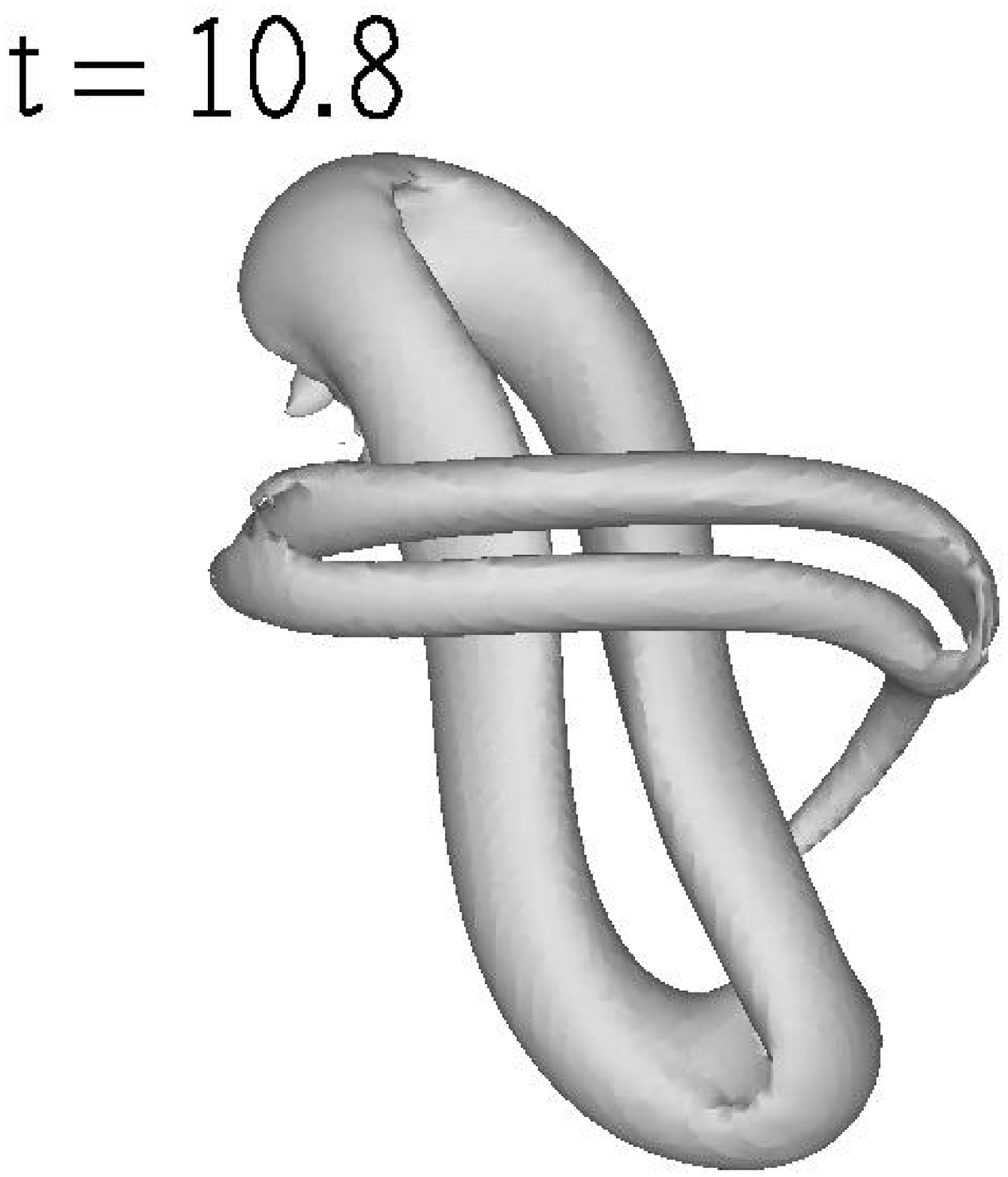}\\ 
\hline 
\end{tabular}
\caption{\label{figure:linked:evol_contours}Linked vortex rings: contours of vorticity; $\omega = 0.025\,\omega_{\text{max}}^{\text{t=0}}$}
\end{minipage}
\hfill 
\begin{minipage}[t]{0.48\linewidth}
\includegraphics[width=\linewidth]{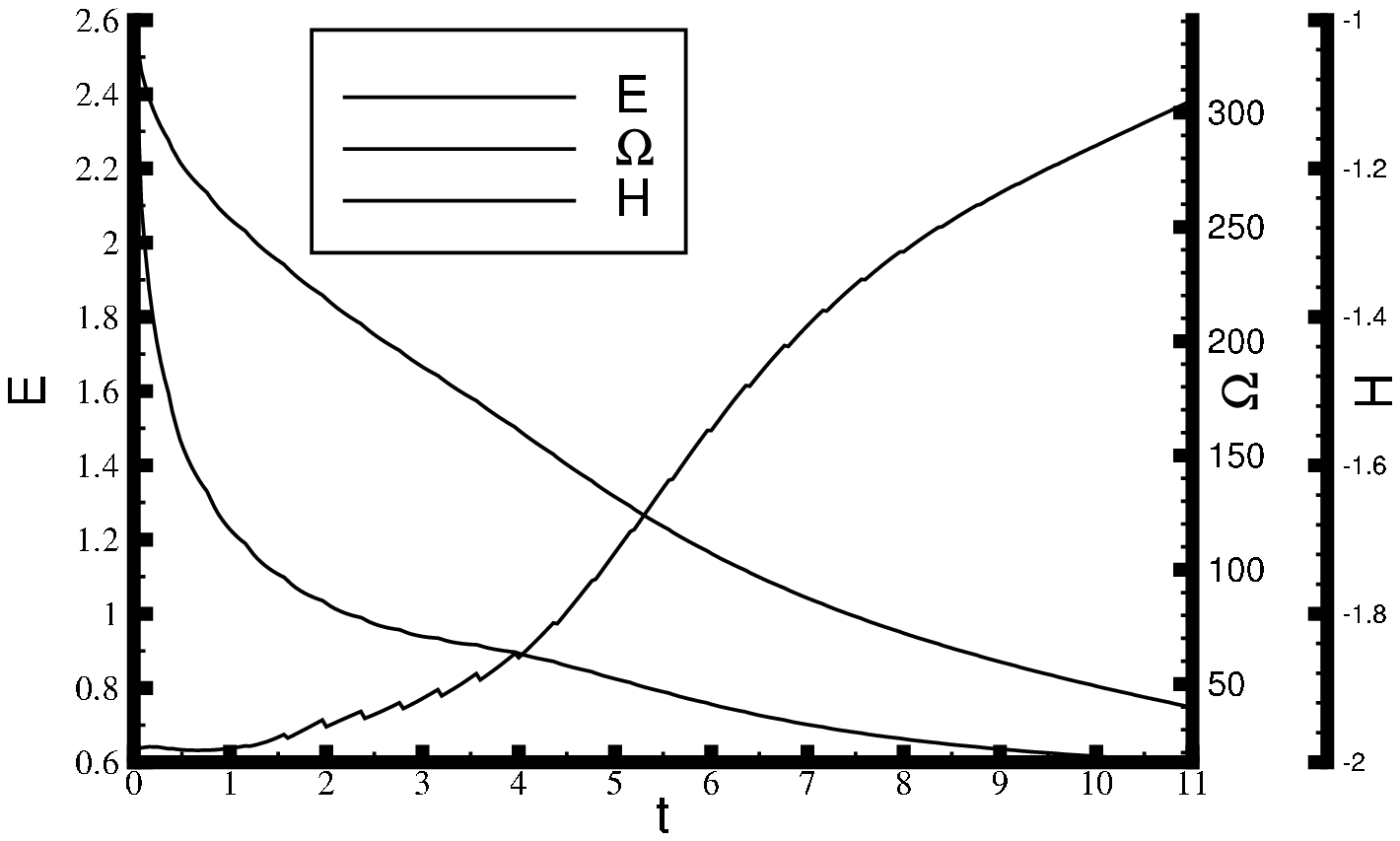}
\caption{\label{figure:linked:evol_stats}Linked vortex rings: kinetic energy, enstrophy and helicity}
\end{minipage}
\end{figure*}
\begin{figure}[h]
\includegraphics[width=0.9\linewidth]{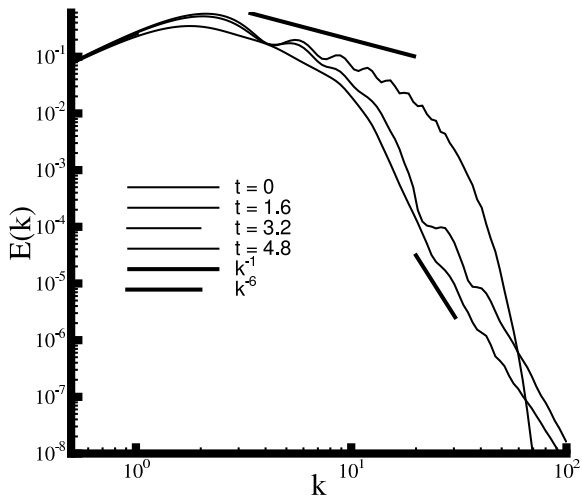}
\caption{\label{figure:linked:spectrum_evolution}Linked vortex rings: evolution of the energy spectrum}
\end{figure}

We study three configurations. In the first case (Fig. \ref{figure:offcoll:evol_contours}, \ref{figure:offcoll:evol_stats} and \ref{figure:offcoll:spectrum_evolution}), the initial rings are placed at a distance 
of $R/4$ apart in the z direction, offset by $R$ along the y axis and they move in opposite directions along the z axis.
The second geometry (Fig. \ref{figure:diffr:evol_contours}, \ref{figure:diffr:evol_stats} and \ref{figure:diffr:spectrum_evolution}), consists of two rings of different radii ($R$ and $R/2$) moving 
in the same direction along the z axis, with the center of the small ring in collision course with the 
circumference of the large one. The small ring has a larger self-induced velocity and catches up with the large ring.
Finally, in case three (Fig. \ref{figure:linked:evol_contours}, \ref{figure:linked:evol_stats} and \ref{figure:linked:spectrum_evolution}), the two rings are linked at $90^o$ a ring going 
through the other in its center. One is moving in the positive z direction; the other, in 
the positive y direction.
\\
All three evolutions have basic common features which will be discussed in the context of the first configuration
for clarity. The spectrum at $t=0$ (Fig.\ref{figure:offcoll:spectrum_evolution}) 
has the characteristic oscillations of the spectrum of isolated vortex rings 
and a cuf-off at the scale of ring core radius $\sigma=0.05$, $k=20$. Our data suggest 
that the reconnection starts around $t=0.6$ when an acceleration of energy decay appears, and 
ends around $t=1.4$. Specifically, as the rings approach each other, they stretch and deform near the 
collision points so that their respective vorticities become locally anti-parallel. The two ends of 
this stretching region eventually become reconnection kinks in which 
(in the absence of singularities), the strong vorticity gradients are smoothed 
out by diffusion. This is also seen in the graphs of the global 
quantities (Fig.\ref{figure:offcoll:evol_stats}) where the beginning of the reconnection 
process corresponds to a hump in the graph of $\Omega$ and to a 
steepening of the slope of $E(t)$ between $t=0.6$ and $t=1.4$. The time length 
of the reconnection is significantly longer than the viscous 
time scale $t_v = \sigma^2/\nu=6.25\,10^{-2}$ and of the same order as the convective 
time $t_c = R^2/\Gamma = 1$. This contradicts the findings of 
 \cite{galcit:schatzle:1987} where it is reported that
the viscous scale is much larger than the reconnection duration. However, 
in  \cite{galcit:schatzle:1987} the
$Re$ number was 1600. The conclusion that 
the duration of reconnection is inversely proportional to the $Re$ number and 
thus to the circulation of the vortices is plausible (also in agreement 
with \cite{jfm:shelley_meiron:1993,jfm:garten_werne:2001}), 
but it is subject to the 
condition in \cite{galcit:schatzle:1987} that the rings are merely
touching themselves rather than colliding.\\ 
After some time ($t=1.6$), we can say that two new rings are formed. 
The pairs of filaments between the reconnection regions are stretched 
further as the new rings move apart from each 
other ($t=1.6$ to $5.6$). Viscous diffusion weakens 
the vorticity magnitude in these structures while the reconnection kinks 
relax in the form of low wavenumber Kelvin waves. These stretched vorticity structures are 
responsible for a continued transfer of energy to the smallest scales until 
these structures are dissipated away. This we conclude by noticing that 
the high wavenumber cut-off of the spectrum becomes 
a non-exponential one (although still very steep) and 
that between $t=0.8$ and $2.4$ there is a significant decay of the energy 
spectrum for $k<20$ but little change for $k>20$.\\ 
It is conceivable that for $Re$ numbers higher than 250 
an intermediate scaling range (that is in between the $k^{-1}$ and $k^{-7}$
regimes) could appear with inertial type of scaling. It is also expected that 
with increasing $Re$ number the hump observed 
in the global enstrophy during the reconnection 
could become more pronounced and (according to the previous discussion) shorter in duration.\\
 The initial helicity is non-zero in the linked configuration. 
As the rings evolve and unlink 
themselves, the corresponding helicity increases as expected and one clearly
notices a slope steepening between $t$=3 
and 5 (Fig.\ref{figure:linked:evol_stats})
when the reconnection action occurs in this case.

In conclusion, we studied three generic
vortex ring configurations and we found that in all
cases the rings reconnect. This suggests that reconnection is a
common phenomenon in vortex filament encounters and perhaps 
also in turbulent flows.
In addition, we observe an intensification of dissipation which is local in time
and could be a mechanism contributing to turbulence intermittency.
A by-product of reconnection is the formation
of stretched structures with anti-parallel vorticity which transfer energy to the smallest scales
where it is rapidly dissipated. Without this energy redistribution in wavenumber space the decay 
of global kinetic energy would have been slower.  This important effect depends directly on
the details of the initial vortex configuration (compare with experiments in \cite{galcit:schatzle:1987}).
The observed intensification of small scale motions hints to an enhancement of small 
scale mixing of passive scalars with $Sc\geq 1$.
The excited Kelvin waves represent a fast mechanism for energy transfer, but the small $Re$ 
number of our calculations is not suitable for understanding their full importance. In particular,
they are confined to low wavenumbers in opposition to the Kelvin waves observed in reconnections in 
quantum fluids \cite{prl:kivotides:2001}. This is because quantum filaments are inviscid and have a very thin core ($\sigma \sim 0.1\,nm$) so that high 
wavenumber Kelvin waves propagate without damping even for rings with 
small circulation.\\
 Besides illuminating important physics,
the present work will guide
future introduction of phenomenological reconnection
models into vortex filament computational methods. In this way,
the applicability of the latter methods will be extended
to flows with complex vorticity configurations.\\
\begin{acknowledgments}
Research partially supported by the Office of Naval Research and the Department of Energy.
\end{acknowledgments}

\bibliography{chatelain}

\end{document}